\documentclass[12pt]{iopart}

\usepackage{iopams}  
\usepackage{amssymb}
\usepackage{mathrsfs}

\def\map#1{\mathcal #1}
\def\d{\mathrm{d}}
\def\<{\langle}\def\>{\rangle}
\def\Tr{\mathrm{Tr}}
\def\:{\hbox{\bf:}}
\def\Cmplx{\mathbb C}
\def\spc#1{\mathscr{#1}}
\def\grp#1{\mathsf{#1}}

\def\Spec{{\rm Spec}}
\def\qed{{$\blacksquare$ \newline}}
\newtheorem{Conj}{Conjecture}
\newtheorem{Proposition}{Proposition}
\newtheorem{Lemma}{Lemma}
\newtheorem{Theo}{Theorem}
\newtheorem{Cor}{Corollary}
\newtheorem{Def}{Definition}

\begin{document}
\title{Optimal asymptotic cloning machines}
\author{Giulio Chiribella} 
\address{Center for Quantum Information, Institute for Interdisciplinary Information Sciences, Tsinghua University\\
Beijing, 100084, China}  
\author{Yuxiang Yang} 
\address{Center for Quantum Information, Institute for Interdisciplinary Information Sciences, Tsinghua University\\
Beijing, 100084, China}

\begin{abstract}
We pose the  question whether the asymptotic equivalence between quantum cloning and quantum state estimation, valid at the single-clone level, still holds when all  clones are examined globally.
We conjecture that the answer is affirmative and present a large amount of evidence supporting our  conjecture, developing techniques to derive optimal asymptotic cloners  and proving their equivalence with estimation in virtually all scenarios considered in the literature. Our analysis covers the case of arbitrary finite sets of states,  arbitrary families of coherent states, arbitrary phase- and multiphase-covariant sets of states, and  two-qubit maximally entangled states. In all these examples we observe that the optimal asymptotic cloners enjoy a universality property, consisting in the fact that scaling of their fidelity does not depend on the specific details of the input states, but only on the number of free parameters needed to specify them.  

 \end{abstract}
\maketitle
\section{Introduction}
Quantum cloning \cite{ScaraniIblisdir05,CerfFiurasek06} and state estimation \cite{Helstrom,Holevo} are two  elemental tasks in  quantum information theory \cite{Werner01,Keyl02,Hayashi}.  They represent  opposite paradigms of information processing:  the coherent processing implemented by quantum machines---which in principle can be reversible---and the incoherent processing based on measurement---which irreversibly turns quantum data  into classical data.  
 Despite the differences, the two tasks of cloning and estimation are deeply related. In particular,   estimation can be used as an intermediate step for cloning \cite{GisinMassar97,BrussEkert98,Werner98}, by using the estimate of the state as an instruction to produce new copies.   This strategy is akin to the way classical copy machines work, by  scanning a document in order to copy it.  However, the key point of quantum cloning is that  the  classical ``estimate-and-copy" approach is typically suboptimal 
\cite{GisinMassar97,BrussEkert98,BrussDivincenzo98,Werner98,CerfIpe00,CerfIblisdir00,namiki1,namiki2,Keyl02,ChiribellaXie13}: in general, the best quantum machine is  a blind device that redistributes the information contained in the input state without making any attempt to read it.

The gap between the performances of cloning and those of  estimation is a basic manifestation of the superiority of quantum information processing over its classical counterpart.  
But  what  happens to this gap in the macroscopic limit where the number of clones becomes large?   This question attracted a considerable amount of interest over the past decade \cite{GisinMassar97,BrussEkert98,KeylWebsite,BaeAcin06,ChiribellaDariano06,Chiribella11}, due to its fundamental relevance and its connection with various security analyses in quantum cryptography.   It is now well known that  the best  quantum machine can be approximated by a machine based on  estimation,  provided that the comparison is made on a small number of clones. Precisely,  Refs.  \cite{ChiribellaDariano06,Chiribella11} showed that the distance between the state of  $k$ clones produced by the quantum machine and the state of $k$  clones produced through estimation goes to zero as $k/M$, where $M$ is the total number of clones.  In other words, cloning is equivalent to state estimation, as long as we restrict our attention to a number of clones that is negligible with respect to the total.

 
 Here we pose the question whether the equivalence between cloning and state estimation continues to hold when one examines all the $M$ clones globally, rather than restricting the attention  to a negligible subset of $k$ clones.     We conjecture that the answer is affirmative:   precisely, we conjecture that the maximum fidelity between the state of the $M$ clones and the state of $M$ ideal copies can be achieved through estimation in the limit $M\to \infty$.  We refer to this  feature as \emph{global equivalence} between asymptotic  cloning and state estimation.  

There are  good reasons to be interested in the global equivalence.   The first reason is conceptual: at the single-copy level, the equivalence with estimation is not a specific feature of cloning, but rather a generic feature shared by all quantum machines  that distribute information in a permutationally invariant fashion \cite{ChiribellaDariano06,Chiribella11}.  Having a global equivalence  is   more interesting, because it highlights something more specific  than  the fact that the output states are invariant under permutations.    
A second, more practical motivation for analyzing the global equivalence comes from the application to quantum money protocols \cite{Wiesner83,Aaronson09,FarhiGosset12,MolinaVidick13}.     In this setting, cloning is the simplest way to attack a protocol, and the global fidelity of the  clones  is the probability that all the counterfeited banknotes pass a test set up by the bank. A global equivalence between cloning and estimation implies that a counterfeiter who aims at producing a large number of copies of the same banknote could do it optimally by scanning the banknote through a measurement, rather than engineering a coherent multipartite interaction that spreads information over the blank copies.    Finally,  establishing whether or not the global equivalence holds is important for the study of quantum benchmarks \cite{HammererWolf05,AdessoChiribella08,OwariPlenio08,CalsamigliaAspachs09,Namiki11,
ChiribellaXie13,ChiribellaAdesso14}, that is, criteria that can be used to certify the advantages of genuine quantum information processing.    If the equivalence did not hold, quantum copy machines would offer an advantage that  persists in the macroscopic limit.  This would provide a  benchmark  detecting quantum features of cloning that are invisible to the single-copy fidelity, and would open the possibility of an experimental demonstration  of quantum advantages in  the macroscopic scenario through the techniques developed by  Ref. \cite{DemartiniSciarrino08,DemartiniSciarrino12}.  

For all these reasons, it is clear that any answer---affirmative or negative---to the question of the global equivalence would have important consequences.   In this paper,  we provide a large amount of evidence in favour of the affirmative,  developing a technique to design optimal asymptotic cloners and showing that the global equivalence holds in all the cloning  problems considered in the literature. Our analysis covers the optimal cloning of  every finite set of states,  of every phase- and multiphase-covariant set of states, of the set of all maximally entangled two-qubit states, and of every family of coherent states---the last category including e.g. arbitrary pure states, spin-coherent states, coherent states of the harmonic oscillator,  squeezed vacuum states and squeezed one-photon states.  
 In many examples, including all the phase/multiphase covariant cloners, we show that the optimal asymptotic cloner  is   \emph{economical}  \cite{NiuGriffiths99}, that is, it is described by a unitary interaction between the $N$ input copies and a set of  $M-N$ blank copies.      This is interesting because economical cloners are the  machines that differ the most from estimation---a difference that can be  tracked back to the difference between the reversible deterministic evolution governed by the Schr\"odinger equation and the irreversible stochastic evolution induced by von Neumann's projection postulate. Proving the equivalence between these two radically different ways of processing information means proving that,  in the macroscopic limit,  the performances of the coherent information processing driven by the Schr\"odinger equation become equal  to the performances of the incoherent processing induced by a measurement.

The paper is organized as follows. In section \ref{sec:preliminary} we formalize our conjecture on the global equivalence between cloning and estimation. The conjecture is then proven  for arbitrary finite sets of states (section \ref{sec:finite}), arbitrary coherent states generated by a group of physical transformations (section \ref{sec:coherent}), multiphase covariant  cloning  (section \ref{sec:qubit}), 
arbitrary phase-covariant cloning (section \ref{sec:arbitraryphase}), and cloning of maximally entangled two-qubit states \ref{sec:entangle}.  In section \ref{sec:distance} we discuss the impossibility to approximate economical cloning channels using estimation.    Finally, the  conclusions are drawn  in section \ref{sec:conclusion}.



 \section{The problem}\label{sec:preliminary}  

\subsection{Optimal cloning of pure states}  
Consider a quantum system with Hilbert space $\spc H$ and a set of unit vectors $\{ |\psi_x\> \}_{x\in\mathsf X}  \subset  \spc H$, representing the  possible input states of the copy machine.   
  The task of optimal quantum cloning  is to convert $N$ identical copies of a state $|\psi_x\>$, chosen at random  with probability $p_x$, into $M$ approximate copies that are as accurate as possible.  The most general  cloning process is described by a quantum channel (completely positive trace-preserving map) $\map C_{N,M}$  transforming density matrices on $\spc H^{\otimes N}$  into density matrices on $\spc H^{\otimes M}$. As a figure of merit for the quality of the copies we consider the \emph{global fidelity}
\begin{equation}\label{generalfid}
F[N \to M]=\sum_{x\in\mathsf X} \,  p_x\,    \Tr\left[  \psi_x^{\otimes M}\map C_{N,M}\left(\psi_x^{\otimes N}\right)\right] \, , 
\end{equation}
where $\psi_x$ denotes the projector  $\psi_x  :  = | \psi_x \>\< \psi_x|$.
When the set $  \{  |\psi_x\>\}_{x\in\mathsf X}$ is continuous, it is understood that the sum has to be replaced by the  integral and the probability $p_x$ is replaced with a probability density $p(x) \d x$. 
The optimal cloner is defined to be the quantum channel that maximizes $F[N\to M]$ and its   fidelity will be denoted by $F_{clon}[N\to M]$.  In some cases it is interesting to consider, instead of the average fidelity of Eq.  (\ref{generalfid}),  the worst case fidelity        
\begin{eqnarray}\label{wc} 
F^* [N \to M]  =  \inf_{x\in\mathsf X}   \Tr\left[  \psi_x^{\otimes M}\map C_{N,M}\left(\psi_x^{\otimes N}\right)\right] \, .
\end{eqnarray}
The advantage of the worst case fidelity is that it does not require us to specify a prior.  
The maximum value of the worst case fidelity over all possible channels will be denoted by $F^*_{clon}  [N\to M]$.  
In general, the cloner that maximizes the worst case fidelity can be different from the cloner that maximizes the average fidelity.  Nevertheless, when the input states are generated by the action of a group and when the prior probabilities are uniform, one can easily prove that the optimal cloners for these two criteria coincide.  

\subsection{Cloning via state estimation} 
The estimation-based  machines are described by  \emph{measure-and-prepare  (MP) channels}, i.e. channels   that  can be realized by measuring the input copies with a  positive operator-valued measure (POVM) $  \{ P_y\}_{y\in\mathsf Y}$  and, conditional on outcome $y$, by re-preparing the system in a state $\rho_y$.  In the case of cloning, the POVM $\{  P_y\}$ acts on the Hilbert space of the $N$ input copies and the states $\{\rho_{y}\}$ are states on the Hilbert space of the $M$ output copies.   
 Averaging over the measurement outcomes,  the output of the MP channel  $\map C_{N,M}$  is given by
 $ \map C_{N,M} (\rho)  = \sum_{y\in\mathsf Y}     \Tr [   P_y \rho]  ~   \rho_y  .$     We denote by $F_{est} [N\to M]$ (respectively, $F^*_{est} [N\to M]$) the maximum of the average (respectively, worst case) fidelity over the set of  MP channels.   Such a  maximum  is known in the literature as \emph{classical fidelity threshold} \cite{HammererWolf05,AdessoChiribella08,OwariPlenio08,CalsamigliaAspachs09,Namiki11,ChiribellaXie13,ChiribellaAdesso14} and can be used as a benchmark for the experimental demonstration of quantum advantages. 
  
\subsection{The problem of the global equivalence }   The key question of this paper is whether the difference between $F_{clon}[N\to  M ] $ and $F_{est}[N\to M]$  (or, alternatively, between  $F^*_{clon}[N\to  M ] $ and $F^*_{est}[N\to M]$) becomes negligible in the limit $M \to \infty$. 
   In the formalization of the problem there is a catch, because in some interesting cases both fidelities tend to zero in the limit $M\to \infty$.  
    For example, the fidelity tends to zero whenever the set of  states to be cloned contains a one-parameter family of ``clock states" $ \{ |\psi_t\>  =   e^{-iHt} |\psi\> \,  ,  t\in\mathbb R  \}$ generated by the action of a Hamiltonian $H$:  in all such cases a non-vanishing fidelity would violate the strong converse of the Standard Quantum Limit for information replication \cite{ChiribellaYang13}, which states that every quantum channel that produces more than $ O(N)$ output copies must necessarily have vanishing fidelity.  In order not to trivialize the question of the global equivalence, it is then important to consider the difference between the two fidelities at the leading order.   For this reason, we formulate our conjecture as follows:  
    \begin{Conj}{\bf (Global equivalence of cloning and state estimation)} For every set of pure states $\{  |\psi_x\> \}$ and prior probabilities $\{p_x\}$, the  global  fidelities of cloning and estimation satisfy the relation
     \begin{eqnarray}\label{conj1}
 \lim_{M \to \infty} \frac{  F_{clon}[N \to M]   -  F_{est}[N  \to M]}{F_{clon}[N \to M]}=0 \qquad \forall N \in\mathbb N \, .
  \end{eqnarray}   
  \end{Conj}
 Of course, the conjecture can be also formulated in terms of the worst-case fidelities, if one prefers not to specify a prior probability distribution. 
When the  equality holds, we say that asymptotic cloning and state estimation are \emph{globally equivalent}, with respect to the average or to the worst-case fidelity. 
In the following sections we will prove the global equivalence in a variety of different settings, exploring in details the structural features of the optimal cloner and  of the optimal MP protocol.

\section{Cloning of finite sets of states}\label{sec:finite} 
We start our investigation from the simplest case, where the set of  states that the machine tries to copy  is finite, say,  $ \{  |\psi_x\> \}_{x\in\mathsf X}$ with  $\mathsf X  =  \{1,\dots, |\mathsf X|\}$.  In this case, the performance of the optimal cloner can be 
upper bounded as follows:
\begin{Proposition}\label{prop:finite}  
For every finite set of states and for every set of probabilities, 
the cloning fidelity satisfies the upper bound
\begin{eqnarray}\label{finitequantum}
F_{clon}[N\to M]  \le  p^{(N)}_{succ} +      O\left(  \eta^{M/2}\right) \, ,
\end{eqnarray}
where $p_{succ}^{(N)}$ is the maximum probability  of correct identification of the input state, given by  
$$  p_{succ}^{(N)}  :   =  \max_{  \{  P_x\}} \,  \sum_{x\in \mathsf X}  \,  p_x \,  \<  \psi_x|^{\otimes N} P_x |\psi_x\>^{\otimes N}  $$ 
(the maximum running over all possible POVMs) and  $\eta  $ is  the maximum pairwise fidelity between two distinct states, given by $ \eta  :=  \max_{x\not = y}  |\<  \psi_x|\psi_y\>|^2$.
\end{Proposition}
Since by definition $\eta  <  1$, the above bound tells us that the cloning fidelity is upper bounded by the probability of success, plus a term that vanishes exponentially fast. The proof of  proposition \ref{prop:finite} is based on the Gram-Schmidt orthogonalization procedure, combined the following bound, which has some interest in its own right:  
\begin{Lemma}\label{lem:gram}
For every set of unit vectors $\{  |\psi_x\>\}_{x\in\mathsf X}$ there exists a set of orthonormal vectors $\{|\gamma_x\>\}$ such that the projectors $\psi_x$ and $\gamma_x$ satisfy the bound
\begin{eqnarray}\label{expbound}
\|  \psi_x  -   \gamma_x \|_{\infty}  \le    ~ \sqrt{  \frac{\alpha^{|\mathsf X|}  \eta }{\alpha -1}  } \,  ,  
\end{eqnarray}
where $  \|  A  \|_{\infty}  =   \sup_{  \|  |\psi\>  \|  =1}   \|  A  |\psi\>  \|  $ is the operator norm and $\alpha =  3 + 2 \sqrt 2$. 
\end{Lemma} 

The proof of the bound is provided in \ref{app:gram}.  Using this result, one can easily prove the bound on the cloning fidelity:
\medskip
 
\noindent {\bf Proof of Proposition \ref{prop:finite}.}  Let $\map C_{N,M}$ be the optimal cloning channel and let $\{     |\gamma_x\>\}$ be the orthonormal states constructed from the states $\{  |\psi_x\>^{\otimes M}\}$ as in lemma \ref{lem:gram}.  With these settings, one has
\begin{eqnarray*}
F_{clon} [N\to M] &  =  \sum_{x}     \, p_x  \,   \Tr  \left[  \psi_x^{\otimes M}    \map C_{N,M}\left (\psi_x^{\otimes N}\right)\right]  \\ 
&  \le     \sum_{x}     \, p_x  \left\{  \Tr  \left[  \gamma_x  \,   \map C_{N,M}\left (\psi_x^{\otimes N}\right)\right]     +     \|  \psi_x^{\otimes M}  -  \gamma_x \|_{\infty}   \right\} \\
&  \le     \sum_{x}     \, p_x  \,   \Tr  \left[  \gamma_x  \,   \map C_{N,M}\left (\psi_x^{\otimes N}\right)\right]   +    \sqrt{  \frac{\alpha^{|\mathsf X|}  \eta^M }{\alpha -1}  }  \\
&  =     \sum_{x}     \, p_x  \,   \Tr  \left[  P_x    \, \psi_x^{\otimes N}\right]   +    \sqrt{  \frac{\alpha^{|\mathsf X|}  \eta^M }{\alpha -1}  }   \, ,
\end{eqnarray*}
where $\{P_x\}$ is the POVM that results from applying the channel $\map C$ and then measuring on  an orthonormal basis that contains the vectors $\{|\gamma_x\>\}$. 
Maximizing  the r.h.s. over all POVMs $\{ P_x\}$ one  obtains the desired bound.  \qed

It is easy to see that the upper bound of Eq. (\ref{finitequantum}) can be achieved by an MP protocol in the asymptotic limit: 
Let $\{  P_y\}_{y\in \mathsf X}$ be the POVM that maximizes the probability of correct identification of the input state.  Then,  consider the naive MP protocol that consists in measuring  the optimal POVM $\{  P_y\}_{y\in \mathsf X}$  and, conditional on outcome $y$,  re-preparing  $M$ copies of the state $|\psi_y\>$.  For this protocol, the fidelity is 
\begin{eqnarray}
\nonumber F  [N\to M]  &=  \sum_{x,y \in\mathsf X} \, p_x  \,  \<  \psi_x|^{\otimes N} P_y |\psi_x\>^{\otimes N}\,  |  \<  \psi_x|   \psi_y\>|^{2 M}    \\
\nonumber &\ge   \sum_{x \in\mathsf X} \, p_x  \,  \<  \psi_x|^{\otimes N} P_x |\psi_x\>^{\otimes N}  \\
 \label{finitemp} &  =  p_{succ}^{(N)} \, .
\end{eqnarray}

Combining Eqs. (\ref{finitequantum}) and (\ref{finitemp})   we can now prove the global equivalence between cloning and estimation.  Indeed, we have
\begin{eqnarray*}
\lim_{M \to \infty}    \frac{  F_{clon}[N \to M]   -  F_{est}[N  \to M]}{F_{clon}[N \to M]}    
  & \le \lim_{M \to \infty}    \frac{    O\left( \eta^{M/2} \right)      } {p_{succ}^{(N)}  +     O\left( \eta^{M/2} \right)        }  =  0 \, .
 \end{eqnarray*} 
In summary,  we demonstrated that, for an arbitrary finite set of states, the following three quantities are asymptotically equal:
\begin{enumerate}
\item the fidelity of the optimal quantum cloner
\item the probability of correct identification of the input state
\item the fidelity of the optimal MP cloner.
\end{enumerate}
Clearly, these equalities  indicate that, asymptotically, the optimal MP protocol is the naive one, which tries to identify the state using the best state discrimination strategy and then re-prepares  $M$ identical copies according to the estimate.   Note that the  same results presented here can be obtained in terms of worst-case fidelities, following the same lines of argument.  




The validity of the global equivalence for arbitrary finite sets of states is  already a strong indication in favour of our conjecture.  In fact, if one believes that infinity is an abstraction and that only finite sets of states play a role in real experiments, then the proof presented here already covers all possible cases of interest.  However, a large number of quantum cloning machines considered in the literature are designed to clone continuous set of states, such as the set of all pure states, or the set of coherent states of the harmonic oscillator.   In these cases, proving the conjecture requires a different argument than the orthogonalization argument used here, because  the upper bound of Eq. (\ref{expbound}) diverges when the number of states $|\mathsf X|$ becomes infinite.    In the following sections we will address the problem in the case of continuous sets of states with symmetry, covering all the examples considered in the literature and providing the optimal asymptotic cloners in a number of new examples.

\section{Cloning of coherent states}\label{sec:coherent}   
The set of all pure states of a finite-dimensional quantum system, the set of spin-coherent states, the sets of coherent, squeezed, and pure Gaussian states  in quantum optics are all examples of a general class of states,  known as \emph{Gilmore-Perelomov  coherent states (GPCS)}  \cite{Gilmore72,Perelomov,Perelomov72}.  Mathematically, GPCS are states of the form $  |\psi_g\> =    U_g  |\psi\>, g\in\grp G$, where $\grp G$ is a Lie group and  $|\psi\>$ is a lowest weight vector for an irreducible representation $U$, i.e. a vector that is annihilated by all the negative roots of the Lie algebra (strictly speaking, the coherent states of the harmonic oscillator are not generated from a lowest weight vector, but we include them in our list because, when it comes to tensor products, they enjoy the same properties of coherent states generated from lowest weight vectors, thus allowing for a unified treatment).    In the following, we will show the validity of the global equivalence for arbitrary GPCS, first giving a general upper bound and then showing how to achieve the bound through estimation.   We will treat separately the case of the average and worst case fidelity, because here the worst-case scenario allows for an exact optimization of the cloner, whose form generalizes the well-known form of the optimal cloner of pure states presented by Werner \cite{Werner98}.  

\subsection{Upper bound on the cloning fidelity}  
Here we  derive  a general upper bound on the optimal cloning fidelity in terms of the probability density of correct identification of the input state, the so-called \emph{likelihood} of the estimation \cite{Helstrom,Holevo}.  The key idea in our argument  is that GPCS can be used to construct a POVM \cite{aliAntoine}, also known as the \emph{coherent state POVM}.   Indeed, thanks to the Schur's lemma one has the identity
\begin{eqnarray}\label{cohpovm}
 \int  \d g ~     \psi_g^{\otimes M}  =    \frac{  P_M}{d_M}  \, ,   
\end{eqnarray}       
where $\d g$ is the Haar measure, $P_M$ is the projector on the subspace of $\spc H^{\otimes M}$ spanned by the states $\left\{ |\psi_g\>^{\otimes M}  \right\}$, and $d_M$ is a suitable normalization constant, given by  
\begin{eqnarray}\label{formaldim}
d_{M}   =   \left( \int  \d g  ~  |\< \psi|  \psi_g\>|^{2M}     \right)^{-1}
\end{eqnarray}  For compact groups,  the Haar measure can be normalized to 1 and the constant $d_M$ is just the trace of  $P_M$, or, equivalently, the dimension of the subspace of $\spc H^{\otimes M}$ spanned by the states $\left\{ |\psi_g\>^{\otimes M}  \right\}$.  For non-compact groups in infinite dimensions, $d_M$ is called formal dimension and in our discussion we assume that $d_M  >0$ for sufficiently large $M$. As a matter of fact, this condition is satisfied in all known cases, including the Weyl-Heisenberg coherent states (where the condition is satisfied for all $M$) and the squeezed vacuum states (where the condition is satisfied for all  $M\ge 3$  \cite{ChiribellaDarianoLorentz06}).    Thanks to Eq. (\ref{cohpovm}), one can define the coherent-state POVM  $\{E_g\}$ via the relation $E_g  :  =  d_M  \psi_g^{\otimes  M}$.

We are now in position to derive our upper bound on the cloning fidelity: 
\begin{Proposition}\label{prop:uppercoh}
For every set of coherent states   $\{  |\psi_g\>\}$   and for every probability density $p(g)$, the cloning fidelity satisfies the bound
\begin{eqnarray}\label{uppercoh}
F_{clon}  [N \to M]  \le   \frac{  p_{succ}^{(N)}}{d_M} \, ,
\end{eqnarray}
where $p_{succ}^{(N)}$ is the maximum probability density of correct identification of the input state, namely 
$$   p_{succ}^{(N)}  :  = \sup_{   \{  P_g\}}  \int  \d g \,  p( g)\,  \< \psi_g|^{\otimes N}    P_g  |\psi_g\>^{\otimes N} \, ,$$
the supremum running over all possible POVMs $\{  P_g\}$.  
\end{Proposition}      
{\bf Proof.}  Let   $  \map C_{N,M}$ be the optimal cloning channel. Then, we have
\begin{eqnarray}
\nonumber F_{clon}[N\to M]  & =  \int  \d g  \, p(g)   \Tr \left[  \psi_g^{\otimes M}  \map C_{N,M} \left ( \psi_g^{\otimes N} \right)  \right] \\
\nonumber  &  =   \frac  1{ d_M}      \int  \d g \,  p(g)   \Tr  \left[   \left(   d_M  \psi_g^{\otimes M}  \right)   \map C_{N,M} \left ( \psi_g^{\otimes N} \right) \right]  \\
\nonumber  &  =   \frac  1{ d_M}      \int  \d g \,  p(g)   \Tr  \left[  P_g   \,  \psi_g^{\otimes N} \right]\\   
\nonumber &  =   \frac  {p_{succ}}{ d_M}  \, ,
\end{eqnarray} 
where $  \{P_g\}$ is the POVM that corresponds to measuring the coherent-state POVM $\{E_g \}$ after the channel $\map C_{N,M}$ and $p_{succ}$ is the probability density of correct identification of the input state with the POVM $\{  P_g\}$, averaged over all possible states.  Optimizing over all possible POVM then gives the desired bound.  \qed 

Note that, in general, there is no guarantee that the upper bound of Eq. (\ref{uppercoh}) is achievable.  In the following paragraphs we will see that the bound is achievable in the limit $M  \to \infty$. Moreover,  we will consider the worst-case scenario, showing that in this case the bound can be achieved for every finite $M$.      

\subsection{An asymptotically optimal MP protocol}
Consider the naive MP protocol, which consists in estimating the input state with the optimal POVM  $\{ P_{g}\}$ and in re-preparing $M$ copies according to the estimate.  In the case of coherent states, it is easy to see that this protocol is asymptotically optimal.    By definition, its fidelity is given by  
\begin{eqnarray*}
F[{N\to M}]  & =    \int  \d g \,  p(g)  \int \d \hat g ~\, |\<  \psi_{\hat g}|\psi_g\>|^{ 2M}  ~  \Tr [  P_{\hat g}   \psi_g^{\otimes N}]  \, .
\end{eqnarray*}

Clearly, in the large $M$ limit the overlap  $|\<  \psi_{\hat g}|\psi_g\>|^{ 2M}$ becomes sharply peaked around the correct value $\hat g=g$, so that in the integral  we can substitute the value  of the slowly varying function $ \Tr [  P_{\hat g}   \psi_g^{\otimes N}]$ with its value at $\hat g  =  g$. With this approximation, we obtain  
\begin{eqnarray*}
F [N\to M]  & \approx    \int  \d g \,  p(g)  \int \d \hat g  ~\, |\<  \psi_{\hat g}|\psi_g\>|^{ 2M}  ~  \Tr [  P_g   \psi_g^{\otimes N}]  \\ 
& =   \frac {1}{d_M}    \int  \d g \,  p(g)   \,  \Tr [  P_g   \psi_g^{\otimes N}]\\
& =   \frac  {p^{(N)}_{succ}}{ d_M}    \, ,  
\end{eqnarray*}
where in the second line we used Eq. (\ref{cohpovm}).    Hence, the naive MP protocol achieves the upper bound of Eq. (\ref{uppercoh}) in the asymptotic limit.    An explicit quantification of the error in the approximation will be provided  for the worst-case fidelity  in paragraph \ref{subsect:cohworst}. 

\subsection{Optimal quantum cloner in the worst-case scenario}
      
In the worst-case scenario, the strong symmetry of the set of coherent states simplifies the optimization of the optimal cloning machine and of the optimal estimation strategy, enabling a complete solution, as shown in the following 
\begin{Proposition}[Optimal cloner of coherent states]
For a given set of coherent states $\{  |\psi_g\>\}$, the optimal worst-case fidelity is   
\begin{eqnarray}\label{gpcsclon}
F_{clon}^*  [N \to M]   =   \frac {  d_N }{d_M}  
\end{eqnarray}   
and is achieved by the optimal cloner 
\begin{eqnarray}
\label{wernerbis}
\map C_{N,M}  (\rho)   =  \frac { d_N}{d_M}  P_M   \left[\rho  \otimes I^{\otimes (M-N)}\right]   P_M \, ,
\end{eqnarray}
where $P_M$ is the projector on the subspace spanned by the vectors $\{|\psi_g\>^{\otimes M}\}$.   
\end{Proposition} 
The  expert reader can recognize in Eq. (\ref{wernerbis}) the same general form of the optimal cloner  defined by Werner for pure states \cite{Werner98}, which is extended here  to coherent states associated to arbitrary groups.    It is also worth noting that also the optimal cloner of coherent states of the harmonic oscillator \cite{CerfIpe00,CerfIblisdir00} can be cast in the form of Eq.  (\ref{wernerbis}).    

\medskip

\noindent {\bf Proof.} 
Following the same steps in the proof of proposition \ref{prop:uppercoh}, one can derive the bound
\begin{eqnarray*}
F_{clon}^*  [N \to M]   \le   \frac {  p^{*\, (N)}_{succ}}{d_M} \, , 
\end{eqnarray*}   
with $   p^{*\, (N)}_{succ}   =    \sup_{  \{P_g\}}   \inf_g   \< \psi_g|^{\otimes N}    P_g  |\psi_g\>^{\otimes N}$, the supremum  running over all possible POVMs .  Now, in the presence of symmetry, the POVM that maximizes the worst case probability  density  $p^{*\, (N)}_{succ}$   is the \emph{maximum likelihood POVM} from Refs. \cite{ChiribellaDariano04,ChiribellaDarianoPerinotti06}.  
In the case of coherent states, the general formula for  the maximum likelihood POVM yields the coherent-state POVM   $\{E_g\}$, given by  $ E_g  = d_N  \psi_g^{\otimes N}$.   Substituting    in the above  bound we then obtain
\begin{eqnarray*}
F^*_{clon}[N\to M]   \le    \frac{  d_N}{d_M}  \, .
\end{eqnarray*}  
Clearly, the channel $\map C_{N,M}$ defined in Eq. (\ref{wernerbis}) achieves the bound.  \qed 

{\bf Remark  (the case of uniform prior).}   Note that, in the case of compact groups, where the Haar measure can be normalized, the  optimal worst-case fidelity is equal to the optimal average fidelity with respect to the uniform prior  $p(g)  \d g   =  \d g$.   Similarly, the worst-case success probability is equal to the optimal average probability with respect to the uniform prior, namely 
$  p^{  *  \, (N)}_{succ}  =  p_{succ}^{(N)}$.   
Moreover, the average density matrix of the target states is $
\rho_{AV}^{(M)}   :=   \int \d g \,   \psi_g^{\otimes M}    =  { P_M}/{d_M} $, 
having used Eq. (\ref{cohpovm}).   Collecting these observations together, we can cast the average fidelity  in the form \begin{eqnarray}\label{primavolta}
F_{clon}  [  N\to M]      =       \left\|   \rho_{AV}^{(M)}   \right\|_{\infty}  ~    p^{(M)}_{succ}  \, .    
\end{eqnarray} 
The expression of Eq. (\ref{primavolta}) will reappear many times in this paper, for sets of states that are not necessarily  coherent states.  In general,  Eq. (\ref{primavolta}) will provide an upper bound on the cloning fidelity, which becomes achievable in the asymptotic limit.  Here, what is specific of coherent states is that Eq. (\ref{cohpovm}) gives the exact value of the cloning fidelity for every $N$ and $M$, not only in the asymptotic limit.

\subsection{Asymptotically optimal MP protocol for compact groups}\label{subsect:cohworst}

In the worst-case scenario, it is possible to prove that the best MP protocol for coherent states is, again, the naive protocol, which consists in   measuring the input copies with the coherent-state POVM  and re-preparing $M$ identical copies  according to the estimate \cite{ChiribellaAdesso13}.     For compact groups, we will now show that  this MP protocol achieves  the optimal quantum fidelity in the asymptotic limit.

The argument is based on a lower bound on the worst-case fidelity of the naive protocol, which reads 
\begin{eqnarray*}
F^*[N\to M] &=  d_N  \int \d g   ~    |\<  \psi  |\psi_g\>|^{2M}       |\<  \psi  |\psi_g\>|^{2N} \\
 &\ge   d_N  \int_{\mathsf S_\epsilon} \d g   ~    |\<  \psi  |\psi_g\>|^{2M}       |\<  \psi  |\psi_g\>|^{2N} \, .
\end{eqnarray*} 
where   $  \mathsf S_\epsilon$ is the set of values of $g$ such that $|  \< \psi|\psi_g\> |^{2M} \ge \epsilon$, for some fixed $\epsilon$.  By definition  $\mathsf S_\epsilon$ satisfies 
$$   \int_{\mathsf S_\epsilon}  \d g \, |  \<   \psi|  \psi_g\>  |^{2  M}    \ge    \int  \d g \, |  \<   \psi|  \psi_g\>  |^{2  M}-\epsilon \,.  $$  
Combining  this fact with the relation $|  \< \psi|\psi_g\> |^{2} \ge \epsilon^{1/M}  \,  \forall g  \in \mathsf S_\epsilon$, it is straightforward to obtain the bound  
\begin{eqnarray}
\nonumber F^*[N\to M]   &\ge   d_N  \, \epsilon^{\frac N M} \,  \int_{\mathsf S_\epsilon} \d g   ~    |\<  \psi  |\psi_g\>|^{2M}   \\
\nonumber &\ge   d_N  \, \epsilon^{\frac N M}   \left( \,  \int  \d g   ~    |\<  \psi  |\psi_g\>|^{2M} -\epsilon\right)\\
\label{fepsilon}&  =  \frac{d_N    \, \epsilon^{\frac N M}   (1-\epsilon\, d_M)}{d_M}     \, ,
\end{eqnarray} 
having used Eq. (\ref{cohpovm}) in the last equality. 
Now, for every compact group, the subspace spanned by the coherent states $\{  |\psi_g\>^{\otimes M}\}$ is contained in the symmetric subspace, and therefore we have the bound 
$d_M     =  O\left( M^{d-1}\right)$. 
Hence, we can set $\epsilon  =  N/(Md_M)$ and obtain the relation 
$\lim_{M\to \infty} \epsilon^{{N}/{M}} =1$,  
which, combined  with Eqs. (\ref{gpcsclon}) and (\ref{fepsilon}), gives 
\begin{eqnarray*}
\lim_{ M\to\infty}  \frac{F^*_{clon} [N\to M]  -  F^*_{est}[N\to M] }{ F^*_{clon} [N\to M]} &    \le  \lim_{M\to \infty}1-\epsilon^{N/M}   (1-\epsilon\, d_M)   =  0 \, .    
\end{eqnarray*}
This provides a quantitative proof of the  equivalence between cloning and state estimation for all families of coherent states associated to compact groups.   
\section{Multiphase-covariant cloning}\label{sec:qubit}

Coherent states are  highly symmetric sets of states.  One may wonder whether the global equivalence between cloning and estimation, shown to be valid for coherent states, remains valid  also for sets of states with a lesser degree of symmetry.       In this Section we  analyze   the case of   multiphase covariant cloning, where the input state is chosen  uniformly at random  among the states of the form
\begin{eqnarray}\label{multiphaseinput}
|\psi_{\vec \theta}  \>   & =     \sqrt {p_0}   |0\>  +     \sum_{j=1}^{d-1} \,  \sqrt{  p_j}  \, e^{  i  \theta_j} |j\>   \, .
 \end{eqnarray}  
Here  $\theta_1,\dots, \theta_{d-1}$ are  phases chosen independently at random in the interval $[-\pi,\pi)$, $\vec \theta$ denotes the vector $\vec  \theta  =  (1, \theta_1,\dots,  \theta_{d-1})$, and $\{  p_j~|~  j  = 0 ,\dots, d -1 \}$ is a fixed probability distribution.   Without loss of generality, we assume that every probability $  p_j$ is strictly larger than zero: obviously,  the case where some probabilities are zero can be reduced to this case by suitably restricting the Hilbert space.  

Multiphase covariant cloning has been studied extensively in the literature \cite{BrussCinchetti00,FanMatsumoto01,FanMatsumoto02,DarianoMacchiavello03,BuscemiDariano05,DurtFiurasek05},   restricting the attention to the ``equatorial" case where  the  probabilities in Eq. (\ref{multiphaseinput}) are uniform.  
However, very little, if anything at all,  is known in the non-uniform case.   In the following, we derive  the optimal asymptotic cloner, proving the global equivalence with state estimation and  discussing the features of the optimal MP protocol.   The multiphase covariant cloning  is a turning point in our analysis,  for three reasons:  First, it is the simplest example where the naive MP protocol is not optimal.  Second, in this example the optimal quantum machine  is economical, and, therefore, fundamentally different from an MP protocol.    Third, the example has an immediate potential of generalization,  providing a direct path to the proof of the global equivalence for  arbitrary phase-covariant cloning and for the cloning of two-qubit maximally entangled states.

\subsection{The symmetries of the problem}
We start by summarizing
 a few facts that will be used in the derivation of the optimal asymptotic cloners.  First of all, the state of the $N$ input copies can be expressed as 
\begin{eqnarray*}
|\psi_{\vec\theta}  \>^{\otimes N}   = \sum_{\vec n   \in   \map P_{N,d}}   \,   e^{  i    \vec n  \cdot  \vec \theta}\,  \sqrt{  p_{ N,  \vec n } } \,  |    N,  \vec  n\>  \, .  
\end{eqnarray*}
where   $\vec n  =  ( n_0,  \dots, n_{d-1})$ is a partition of $N$ into $d$ non-negative integers,   $\map P_{N,d}$ denotes the set of such partitions,  $  |N, \vec n\>$  is the unit vector that is obtained  by projecting the state $  |0\>^{\otimes n_0}  |1\>^{  \otimes  n_2}  \cdots  |d-1\>^{\otimes n_{d-1}}$ in the symmetric subspace, and   $p_{N,\vec n}$ is the multinomial distribution
\begin{eqnarray*}
p_{  N, \vec n}  =     \frac{N!}{   n_0!  \dots n_{d-1}!  } ~    p_0^{n_0}  \dots  p_{d-1}^{n_{d-1}}  \, .      
\end{eqnarray*}   

For the uniform probability distribution over the states $  |\psi_{\vec \theta}\>$ we will use the notation
$$  \frac{\d \vec \theta}{(2\pi)^{d-1}}    :=      \frac{\d \theta_1  \cdots  \d \theta_{d-1} }{(2\pi)^{d-1}} \, .$$
Since the input state $  |\psi_{\vec \theta}\>^{\otimes N}$ is chosen with uniform probability, the optimal cloning channel can be chosen without loss of generality to be covariant, that is, satisfying the property  
\begin{eqnarray}\label{multiphasecov}    U_{\vec \theta}^{\otimes M}   \map C_{N,M} (\rho)    U_{\vec \theta}^{\dag \otimes M}    =   \map C_{N,M} \left(  U_{\vec \theta}^{\otimes N}   \rho    U_{\vec \theta}^{\dag \otimes N}  \right) \, ,
\end{eqnarray}  for every density matrix $\rho$ and  for every vector $\vec \theta$.  
Covariant channels are known to be optimal also in the worst case scenario, and  the maximum of the average fidelity coincides with the average of the worst case fidelity.     For this reason, in the following we will carry out the analysis in the average scenario, taking for granted that all the results can be translated immediately into worst-case results.

\subsection{Upper bound on the cloning fidelity}
In order to find the optimal asymptotic cloner, we  first  derive an upper bound on the global fidelity:  
 \begin{Proposition}\label{prop:upperphase}
For every phase-covariant set of qudit states  $\{  |\psi_{\vec\theta}\>\}$, the cloning fidelity satisfies the bound
\begin{eqnarray}\label{upperphase} 
F_{clon}  [N \to M]  \le    \left\|   \rho_{AV}^{(M)}  \right\|_{\infty}  \,   p_{succ}^{(N)}       \, ,
\end{eqnarray}
where   $\left  \|    \rho^{(M)}_{AV} \right\|_{\infty}$ is the maximum eigenvalue of the average target state 
$  \rho_{AV}^{(M)}  :   =  \int    {  \d \vec \theta }/{(2\pi)^{d-1}}     \,   \psi_{\vec \theta}^{\otimes M}$ 
and $p_{succ}^{(N)}$ is the maximum probability density of correct identification of the input state, namely 
$  p_{succ}^{(N)}  :  = \max_{   \{  P_{\vec \theta}\}}  \int        {  \d \vec \theta }/{(2\pi)^{d-1}}      \,       \< \psi_{\vec \theta}|^{\otimes N}    P_{\vec \theta}  \, |\psi_{\vec \theta}\>^{\otimes N} $,
the maximum running over all possible POVMs $\{  P_{\vec \theta}\}$.
\end{Proposition}      
The proof of the upper bound is presented in  \ref{app:proofuppermultiphase} and  is based on two ingredients:   the first ingredient is an upper bound on  the fidelity derived from the optimization over all possible covariant cloners, the second ingredient is a translation of the upper bound in terms of the quantities $\|  \rho_{AV}^{(M)}\|_\infty$ and $p_{succ}^{(N)}$.  
Since this translation will be used several times in the following, we discuss it explicitly here:  
Regarding $\|  \rho_{AV}^{(M)}\|_\infty$,  this can be easily computed using the Schur's lemma, which gives 
$ \rho_{  AV}^{(M)}   =  \sum_{\vec m \in  \map P_{M,d}}      \, p_{M,\vec m} \,  |M, \vec m\>\<  M, \vec m| $,
and, therefore  
\begin{eqnarray}\label{rhoavmax}
\left\|  \rho_{AV}^{(M)}\right\|_\infty  =   \max\{ p_{M,\vec m} ~|~ \vec m\in\map P_{M,d} \} \, .  
\end{eqnarray}
Regarding $p_{succ}^{(N)}$,  we use the general expression of the optimal POVM  for sets of pure states with group symmetry, provided in Refs. \cite{ChiribellaDariano04,ChiribellaDarianoPerinotti06}. In the case of multiphase covariant states, the general recipe yields the optimal POVM
\begin{eqnarray}
\label{multitheta}
P_{\theta} & =   |\eta_{\vec \theta}\>\< \eta_{\vec \theta}| \, , \qquad  |\eta_{\vec  \theta}  \>   & := |0\>  +   e^{i \theta_1}  |1\>  +  \dots  +  e^{i\theta_{d-1}} |  d-1\>   \, 
\end{eqnarray}  
and the maximum probability density  $p_{succ}^{(N)}$ is given by 
\begin{eqnarray}\label{psuccmultiphase}  p_{succ}^{(N)}  =    \<  \eta_{\vec \theta}|   \psi_{\vec \theta}^{\otimes M}  |\eta_{\vec \theta}\>     = \left(  \sum_{\vec n\in\map P_{N,\vec n}}  \sqrt{  p_{N,\vec n}}\right)^2  \qquad \forall \vec \theta   \, .
\end{eqnarray}

In the  next two  paragraphs we will show two, radically different cloning strategies that match the upper bound of Eq. (\ref{upperphase}) in the asymptotic limit. The first strategy  consists in using an economical quantum channel, which coherently encodes the  $N$ input copies into the space of $M$ systems.  The second strategy consists in using a suitable MP protocol---this time, however, not the naive one. 

\subsection{An asymptotically optimal economical cloner}
Let us start by showing a quantum strategy that achieves the upper bound of Eq. (\ref{upperphase}) in the asymptotic limit.  Consider  the economical  channel $\map C_{N,M}$ defined by 
\begin{equation}
\label{ecomulti}  \map C_{N,M} (\rho)    =  V_{N,M} \rho V_{N,M}^\dag \qquad   
 V_{N,M}  =  \sum_{\vec n  \in  \map P_{N,d} }    | M, \vec n  - \vec n_* +  \vec m_*     \>\<  N,  \vec n |   \, . 
\end{equation}  
 where  $\vec n_*$ (respectively,  $\vec m_*$) is the partition that maximizes the multinomial probability $  p_{N, \vec n}$  (respectively, $p_{M,\vec m}$).  The cloning fidelity of channel  of this channel is
\begin{eqnarray}
\nonumber F[N\to M]  &  =    \int    \frac{  \d \vec \theta }{(2\pi)^{d-1}}     \, \left|  \<\psi_{\vec \theta}|^{\otimes M}    V_{N,M}   |\psi_{\vec \theta}\>^{\otimes N}\right|^{2}\\
\label{eq:qubitclone} &=\left(\sum_{\vec n\in\map P_{N,d}}  \sqrt{p_{N,\vec n} \, p_{M,  \vec n  - \vec  n_* + \vec m_*}}\right)^2.
\end{eqnarray}
Clearly, when $M$ is large compared to $N$, the multinomial  probability $p_{M, \vec  n- \vec  n_*  +\vec m_*}$   is approximately constant in an interval   of size $O(N)$ centred around $\vec m_*$, up to an error of order $N^2/M$.  Hence,   the fidelity can be approximated as
\begin{eqnarray}
\nonumber F[N \to M ]  & = p_{M,\vec m_*}\left(\sum_{\vec n\in\map P_{N,d}}\sqrt{p_{N,\vec n}}\right)^2   \, \left[ 1  + O  \left(   \frac {N^2} M\right) \right] \\
\nonumber &\equiv  \left\|  \rho_{AV}^{(M)}  \right\|_{\infty}  \, p_{succ}^{(N)}   \, \left[ 1  + O  \left(   \frac {N^2} M\right) \right] \, ,
\end{eqnarray}
having used Eqs.  (\ref{rhoavmax}) and (\ref{psuccmultiphase}).
Comparing this value with the upper bound of Eq. (\ref{upperphase}), we conclude that our cloner is asymptotically optimal.  

\medskip 

\noindent {\bf Remark (large $N$ asymptotics).}   The fidelity of the
 economical cloner  has a  simple and intriguing expression when both $M$ and $N$ are large.  In this case, the multinomial distribution  $p_{N,\vec n}$ can be approximated as 
 \begin{equation}\label{gaussianformulti}  p_{N,\vec n}   =  G_{N,\vec n}   \left  [   1+  O\left( \frac 1  {  \sqrt{N^{1-\delta}}} \right) \right]  +  \eta_N \, ,
 \end{equation}
 where   $G_{N,\vec n}$ is the Gaussian distribution
$$G_{N,\vec n} :  = \frac{  \exp  \left  [   -  \sum_{j=1}^d  \frac{    (  n_j  -   N  p_j   )^2} {  2  N p_j }  \right]}{  \sqrt{(2\pi  N)^{d-1}     ~ p_0  p_1  \dots p_{d-1}  }}   \, ,  $$
 $\delta >  0$ is arbitrary, and $\eta_N$ is an error term vanishing  faster than the inverse of every polynomial.  The Gaussian can be expressed conveniently as 
\begin{eqnarray}\label{gauss}G_{N,\vec n} :  = \sqrt  {   \det  \left   (  \frac{  A}{ 2\pi  N}  \right) }       \exp  \left  [   -   \frac{    \vec x^T  A   \vec x} {  2  N }  \right] \, ,  
\end{eqnarray}  
where $\vec x  =  (  x_j)_{j=1}^{d-1}$  is the vector of the independent variables  $x_j  = n_j  -  N  p_j $, and  $A$ is the positive non-singular matrix  with entries
$ A_{jk}   :=  1 /  p_j  \,    \delta_{jk}   +  1/ {p_0}$.  
 Replacing  the summation in Eq. (\ref{eq:qubitclone}) with a Gaussian integral,  one gets
\begin{equation*}
F_{clon} [  N  \to M]  =   \left(  \frac{  \sqrt{  4 M N} }{M+N}  \right)^{d-1} \,  \left[ 1  +    O  \left(  \frac 1  {\sqrt{N^{ 1-\delta}}}  \right) \right]  \, ,  
\end{equation*} 
for arbitrary $\delta>  0$ .  What is intriguing here is that  the asymptotic fidelity depends only on the dimension of the Hilbert space, and not on the actual values of the probabilities  $\{  p_j\}_{j=0}^{d-1}$ (as long as they are  non-zero). Apparently, the asymptotic limit washes out many of the details of the family of input states, so that the fidelity depends only on a very coarse-grained information, namely, the number of parameters needed to describe the states.  Note that, when some probabilities are zero, the asymptotic formula for the optimal fidelity still holds, provided that  one replaces $d$ with the number of non-zero probabilities. 
 
\subsection{An asymptotically optimal MP protocol}\label{subsec:qubitmeasprep}
We now show that the optimal cloning fidelity can be achieved asymptotically by a suitable MP protocol.  For the first time, here we need to consider an MP protocol that is different from the naive one: as we  anticipated in the introduction, the naive MP protocol  is asymptotically suboptimal.   Intuitively, the reason why re-preparing $M$ identical copies is suboptimal is that the state $    |\psi_{\vec \theta}\>^{\otimes M}$ is very sensitive to small variations of $\vec \theta$.  Hence, a small error in the estimate of the input state   
 results in the re-preparation of a state that is almost orthogonal to  the target state.    
In order to avoid this drawback, one idea is to   re-prepare a number of copies $K$ that is smaller than $M$, but still sufficiently large to mimic, in some way, the target state of $M$ copies.    Driven by the above intuition, we consider  an MP protocol of the following special form:  
\begin{enumerate}
\item   estimate $\theta$ from the $N$ input copies, using the optimal POVM of Eq. (\ref{multitheta})
\item   prepare $K  =  \left \lceil M^{1-\epsilon}  \right \rceil$   copies of the estimated state, for some  $\epsilon   \in  (0,1)$
\item produce $M$ clones via the economical $K$-to-$M$ cloning of Eq. (\ref{ecomulti}). 
\end{enumerate}
Let us see that the protocol achieves the optimal cloning fidelity for every $\epsilon  >  0$.  First of all,  note that the fidelity can be written as  
\begin{equation*}
F_K [  N \to M]=\int         \frac{  \d \vec \tau }{(2\pi)^{d-1}}       
 ~     p_N (\vec \tau |0)   \,   f_{K,M}  (\vec  \tau) \,  ,  
\end{equation*}
with $  \vec \tau:  =  (1,\tau_1,\dots, \tau_{d-1})$,  
 $p_N (\vec \tau|0)    :=   \< \eta_{\vec \tau} |   \,   \psi^{\otimes N}\,  |\eta_{\vec \tau}\>$, and
 $f_{K,M} (\vec \tau\,)    : =   \left|   \<  \psi|^{\otimes M}    V_{   K,M}    |\psi_{\vec \tau}\>^{\otimes K}     \right|^2 $.
For  large $M$ and  $  K$, the Gaussian approximation to the multinomial distribution gives  
$$     f_{M,K}  (\vec \tau\,)   =    \left(\frac{2  \sqrt{MK}}{M+K}   \right)^{d-1} \, \exp \left  [-   \frac  { 2 MK}  {M+K}   ~     \vec \tau^{\, T}  A^{-1}   \vec \tau ~ \right]  \, \left[ 1  + O  \left(   \frac 1{\sqrt{ K^{1-\delta}}  }\right) \right]   +\eta_K  \, ,$$
with arbitrary $\delta>0$ and $\eta_K$ vanishing faster than the inverse of every polynomial.  
Now, when $M$ and $K$ are large compared to $N$, the  Gaussian $f_{M,K}  (\vec \tau\, )$ decays rapidly when $\vec \tau$ moves away from the peak $\vec \tau =  (0,\dots, 0)$. Hence, the slowly varying function $p_N(\vec \tau|0)$ can be treated as a constant in the integral.    Using this fact, we obtain  
\begin{eqnarray}
\nonumber F_K [  N \to M]  &=      p_N  (0|0 )   \,           \left[   \int      \frac{  \d \vec \tau }{(2\pi)^{d-1}}      ~    f_{K,M}  (\vec \tau)  \right]  \, \left[ 1  +   O  \left(   \frac 1{\sqrt{ K}  }+    \frac {N} {K}\right) \right]    +  \eta_K \\
\label{fknm}&  
=     p_N (0|0)  \,   
 \sqrt{ \det\left[   \frac { A }{2\pi  (M  + K  )}   \right]} \left[ 1  +   O  \left(   \frac 1{\sqrt{ K^{1-\delta}}  }+    \frac {N} {K}\right) \right]  \qquad
\end{eqnarray}
Now, by definition we have 
$p_N(0|0)    =  \< \eta |   \psi^{\otimes N}  |\eta\>    \equiv   p_{succ}^{(N)} $   
[cf. Eq. (\ref{psuccmultiphase})] and
\begin{eqnarray*} 
    \sqrt{ \det\left[   \frac { A }{2\pi  (M  + K  )}   \right]} 
          =   \left\|  \rho^{(M)}_{AV} \right\|_{\infty}   \,  \left[1   +    O  \left(   \frac K M   +   \frac1{  \sqrt{M^{1-\delta}}} \right) \right]\,  .
\end{eqnarray*}
Hence,  we obtained
\begin{eqnarray*}
 F_K [N\to M ]& =     p^{(N)}_{succ}  ~  \left \| \rho_{AV}^{(M)}  \right \|_{\infty}   \left[ 1  +   O  \left(   \frac 1{\sqrt{ K^{1-\delta}}  }+    \frac {N} K  +  \frac K M\right) \right]    \\
\end{eqnarray*}
Comparing with the bound of Eq. (\ref{upperphase}), it is clear that our MP protocol is asymptotically optimal. In summary, this establishes the global equivalence for arbitrary multiphase-covariant states.  




\subsection{Suboptimality of the naive MP protocol}

Here we prove that the naive MP protocol is suboptimal, even in the asymptotic limit. 
This protocol has the same  structure of the optimal protocol that we introduced in the previous paragraph, except for the fact that in the naive protocol one  has $K  = M$ instead of $K= \left \lceil  M^{1-\epsilon}  \right\rceil$.  Making this substitution, we can compute the cloning fidelity  using Eq. (\ref{fknm}), which now gives 
\begin{eqnarray*}
F_{K=  M}  [N \to M]  & =           p_N (0|0)  \,   \sqrt{ \det\left[   \frac { A }{4\pi  M }   \right]}        \left[ 1  +   O  \left(   \frac 1{\sqrt{ M^{1-\delta}}  }+    \frac {N} M \right) \right] \\
  &= \frac{F_{clon} [N \to M]}{\sqrt{2^{{d-1}}}}       \left[ 1  +   O  \left(   \frac 1{\sqrt{ M^{1-\delta}}  }+    \frac {N^2} M  \right) \right]\, .
  \end{eqnarray*}
Clearly, this relation shows  that re-preparing $M$ identical copies is a  suboptimal strategy,  even in the asymptotic limit. Moreover,  the gap between the fidelity of the optimal cloner and the fidelity of the naive MP protocol increases exponentially fast with the dimension $d$.   Finally, note that the ratio between the fidelity of the naive MP protocol and the optimal cloning fidelity is independent of the specific values of the probabilities $\{p_j\}$ that define the set of input states  in Eq. (\ref{multiphaseinput}), as long as the probabilities are non-zero.    Again, we see that the asymptotic limit washes away the information about the set of input states, so that the ratio of the fidelities depends only on the number of parameters. 


\section{General phase covariant cloning}\label{sec:arbitraryphase}
The arguments  devised for phase covariant qudit cloning can be also generalized to arbitrary instances of phase covariant cloning, where the input state is chosen uniformly at random among the states of the form  
\begin{eqnarray}\label{phaseinput}
 |\psi_\theta\>   =   U_\theta  |\psi\>    \qquad  U_{\theta}  =   e^{-  i \theta  H}   \, , \theta  \in [-\pi,\pi)
\end{eqnarray} 
where $|\psi\>\in \spc H  $ is a fixed  state and $H$  is some generator with integer spectrum.    One can think of these states as \emph{clock states}, generated from an initial state through the time evolution governed by the Schr\"odinger equation  with Hamiltonian $H$ \cite{ChiribellaYang13}.
The optimal phase covariant cloner is not  known in general. Our strategy will be to first prove an upper bound on the fidelity and then to exhibit cloners that achieve the bound.  As in the previous section, we will show two ways to achieve the bound: with an economical cloner and with an MP protocol.

\subsection{Decomposition of the input state}

Let us denote by $\Spec (H)$ the (integer) spectrum of $H$ and expand the input state $|\psi\>$ as 
\begin{eqnarray*}
|\psi\>  =  \sum_{  E  \in  \Spec (H)}  \,  \sqrt{p_E}    \,  |E\>  \, .
\end{eqnarray*}
where $|E\>$ is an eigenvector of $H$ for the eigenvalue $E$ and $p_E\ge 0$ is the probability that a measurement on the eigenspaces of $H$ gives outcome $E$.  Without loss of generality, we choose $H$ so that
\begin{eqnarray*}
\<  \psi|  H  |\psi\>    =\sum_{E\in  \Spec (H)} \,  p_E  \, E=  0 \, .     
\end{eqnarray*}
Consider now the $N$-copy Hamiltonian, given by $  H^{(N)}  =  \sum_{n=1}^N  H_n$ where $H_n$  is the Hamiltonian $H$ acting on the $n$-th copy.  The $N$-copy state can be expanded as 
\begin{eqnarray*}
|\psi\>^{\otimes N}  =  \sum_{  E  \in  \Spec \left(H^{(N)}\right)}  \,  \sqrt{p_{N,E}}    \,  |N,E\>  
\end{eqnarray*}
where   $|N,E\>$  is an eigenvector of $H^{(N)}$ for the eigenvalue $E$ and $p_{N,E}$ is a probability.  For large $N$ the probability distribution $p_{N,E}$ converges to a Gaussian  centred around the $E=  0$ and with variance proportional to $N$, denoted by
\begin{eqnarray}\label{ge}
g_{N,E}   =  \frac 1  {\sqrt{2 \pi  \<  H^2\>  N}}  ~   \exp\left[  \frac{-  E^2}{2  \<  H^2  \> N}\right]    \,  ,
\end{eqnarray} 
where $ \<  H^2 \>  =  \<  \psi|  H^2  |\psi\>$.  
More precisely, we can put the Gaussian approximation in the form  
$$    p_{ N,E}   =     c \,    g_{N,E}  \left[ 1+  O\left(  \frac 1  {\sqrt{N^{1-\delta}}}\right) \right]  +  \eta_N  \, ,$$
where  $c>0$ is a proportionality constant taking into account the conversion from a finite probability distribution to a continuous one,  $\delta  >  0$ is arbitrary and $\eta_N$ is an error vanishing faster than the inverse of every polynomial.    This form of the Gaussian approximation can be obtained by expressing the energy in terms of partitions of $N$ into $|\Spec (H)|$ non-negative integers   \cite{ChiribellaYang13}, applying the Gaussian approximation of Eq. (\ref{gaussianformulti}) to the multinomial distribution of the partitions, and finally taking the marginal over all partitions that give the same energy.   
  


\subsection{Upper bound on the cloning fidelity}

As one can easily expect, the upper bound on the fidelity derived for multiphase covariant cloners can be  generalized to arbitrary phase covariant sets  in the following way: 
 \begin{Proposition}\label{prop:upperclock}
For every phase covariant set of states  $\{  |\psi_\theta\>\}_{\theta  \in [-\pi,\pi)}$, the cloning fidelity satisfies the bound
\begin{eqnarray}\label{upperclock} 
F_{clon}  [N \to M]  \le    \left\|   \rho_{AV}^{(M)}  \right\|_{\infty}  \,   p_{succ}^{(N)}       \, ,
\end{eqnarray}
where   $\left  \|    \rho^{(M)}_{AV} \right\|_{\infty}$ is the maximum eigenvalue of the average target state 
$   \rho_{AV}^{(M)}  :   =  \int  {  \d \theta}/{2\pi}      \,   \psi_{\theta}^{\otimes M}  $ 
and $p_{succ}^{(N)}$ is the maximum probability density of correct identification of the input state, namely 
$   p_{succ}^{(N)}  :  = \max_{   \{  P_\theta\}}  \int  {\d \theta}/{2\pi}  \,  \< \psi_\theta|^{\otimes N}    P_\theta  |\psi_\theta\>^{\otimes N} \, ,$
the maximum running over all possible POVMs $\{  P_\theta\}$. 
\end{Proposition}      
The steps of the proof are the same of those in the proof of Eq. (\ref{upperphase}) and are omitted since they do not provide additional insight.    

\subsection{Asymptotically optimal economical cloner }
Following the path delineated for multiphase states, it is easy to come up with an economical cloner that saturates the upper bound  of Eq. (\ref{upperclock}) for large $M$:  just choose the economical cloner  $\map C_{N,M}$ defined by 
\begin{equation}\label{isoE}
 \map C_{N,M}  (\rho)  =  V_{N,M}  \rho  V_{N,M}^\dag \, ,  \qquad V_{N,M}    :  = \sum_{  E  \in  \Spec \left(  H^{(N)}\right)}   |  M,  E  +  E_0\>\<  N,E|  \, ,  \quad
\end{equation} 
where $  E_0 $ is the eigenvalue of $H^{(M)}  -  H^{(N)} $ with minimum modulus. This cloner   has fidelity   
\begin{eqnarray}
\label{basta}F  [ N \to N]  &  =  \left( \sum_{E   \in     \Spec \left(  H^{(N)}\right)}        \sqrt{  p_{M, E  + E_0} \,   p_{N,  E}} \right)^2 \, .
\end{eqnarray} 
For large $M$,  we can use the fact that $p_{M,E}$  is almost constant in $\Spec \left(  H^{(N)}\right) $.   In this way, we obtain 
\begin{eqnarray*}
F  [ N \to N]  
\nonumber    &    =  p_{M, E_*} \,   \left( \sum_{E   \in     \Spec \left(  H^{(N)}\right)}           \sqrt{     p_{N,  E}} \right)^2    \left[  1+  O  \left(  \frac {N^2}M  \right) \right]
\end{eqnarray*}
where $E_*$ is the value that maximizes $p_{M,E}$.  
Now, by definition we have $  p_{  M,  E_*}  =  \left\|   \rho^{(M)}_{AV} \right \|_{\infty}  $.  On the other hand, the quantity $\left(  \sum_{E} \sqrt{  p_{N,E}   }  \right)^2 $  can be identified with $p_{succ}^{(N)}$, using Holevo's classic result on phase estimation \cite{Holevo}. 
In conclusion, we obtained   
$$  F  [ N \to M]     =      \left\|   \rho^{(M)}_{AV} \right \|_{\infty} ~    p_{succ}^{(N)}    \,   \left[  1+  O  \left(  \frac {N^2}M  \right) \right]  \, ,$$  proving the achievability of the bound in Eq. (\ref{upperclock}).


\medskip 

\noindent {\bf Remark  (large $N$ asymptotics). }   When both $M$ and $N$ are large, the Gaussian approximation provides the following  simple expression for the fidelity of our economical cloner
\begin{equation*}
F_{clon} [  N  \to M] =  \frac{  \sqrt{4   M N} }{M+N}    \,  \left[ 1  +    O  \left(  \frac 1  {\sqrt{N^{  1-\delta}}}  \right) \right]  \, ,
\end{equation*}
with arbitrary $\delta >  0$.     Note that the value of the fidelity is independent of the Hamiltonian $H$ and of the state $  |\psi\>$, as long as the variance of $H$   is non-zero  on $|\psi\>$.   Moreover,  note  that the fidelity approaches to 1 whenever  the number of extra-copies $M-N$ is negligible compared to $N$, whereas it approaches 0 whenever $N$ is negligible compared to $M$.  This fact is a concrete illustration of the Standard Quantum Limit for cloning established in Ref. \cite{ChiribellaYang13}, which can be derived from general arguments about quantum metrology.

\subsection{Asymptotically optimal MP protocol}
Here we show that the upper bound of Eq. (\ref{upperclock}) can be asymptotically achieved by an MP protocol.  
In order to reach the bound, we follow the same prescription used in subsection \ref{subsec:qubitmeasprep} for multiphase-covariant cloning:  \emph{i)}  estimate the state using the optimal POVM,  \emph{  ii)} if the estimate is $\tau$, then prepare $K$  copies of $  |\psi_{\tau}  \>$, and  \emph{iii)} generate $M$ approximate copies using  the optimal  $K$-to-$M$ economical cloner.   Again, we show that choosing $K  =  \left \lceil M^{1-\epsilon}  \right \rceil$ for some $\epsilon   \in  (0,1)$ allows one to achieve the maximum cloning fidelity in the limit $M  \to \infty$. For a given $K$, the fidelity of our protocol can be written as  
\begin{equation*}
F_K [  N \to M]=\int_{-\pi}^\pi \frac{\d \tau}{2\pi} ~     p_N (\tau|0)   \,   f_{K,M}  (\tau) \,  ,  
\end{equation*}
with 
 $p_N (\tau|0)    :=   \< \eta_{\tau} |   \,   \psi^{\otimes N}\,  |\eta_{\tau}\>$ and 
 $f_{K,M} (\tau)   : =   \left|   \<  \psi|^{\otimes M}    V_{   K,M}    |\psi_{\tau}\>^{\otimes K}     \right|^2 $.
For  large $M$ and  $  K$, the Gaussian approximation gives  
$$     f_{K,M}  (\tau)   =   \frac{2  \sqrt{MK}}{M+K}    \, \exp \left  [-   \frac  { 2 MK \< H^2\>  \tau^2}  {(M+K) c^2} \right]     \left[  1  +  O \left(   \frac 1  {\sqrt{  K^{1-\delta}}} \right)\right] +\eta_K \, .$$
When $M$ and $K$ are large compared to $N$, the  Gaussian $f_{M,K}  (\tau)$ decays rapidly when $\tau$ moves away from the peak $\tau =  0$, and, therefore, the slowly varying function $p_N(\tau)$ can be treated as a constant in the integral.   Using this fact, we obtain  
\begin{eqnarray*}
\nonumber F_K [  N \to M]  &  =    p_N  (0|0 )   \,           \left[   \int_{-\infty}^{\infty} \frac{\d \tau}{2\pi} ~    f_{K,M}  (\tau)  \right]   \,     \left[  1  +  O \left(   \frac 1  {\sqrt{  K^{1-\delta}}}    +    \frac {N}K \right)\right]  +\eta_K \\
\nonumber
&   =     p_N (0|0)  \,    \sqrt{\frac { c^2 }{2\pi   \< H^2\>   (M  + K  )}}\left[  1  +  O \left(   \frac 1  {\sqrt{  K^{1-\delta}}}    +    \frac {N}K \right)\right] \\
&  =  p^{(N)}_{succ}  \,  \left\|  \rho^{(M)}_{AV} \right\|_{\infty}  \, \left[ 1  +   O  \left(   \frac 1{\sqrt{ K^{1-\delta}}  }+    \frac {N} K  +  \frac K M\right) \right]  ,  
\end{eqnarray*}
having used the fact that, by construction of the protocol, $p_N(0|0)   =  p_{succ}^{(N)} $  and, by direct inspection,   $ \|  \rho_{AV}^{(M)}\|_\infty  =  p_{M,E_*}   =  c  g_{M,0} [  1 +   O (    1/\sqrt{M^{1-\delta}})]  $. 
In conclusion, our MP protocol achieves asymptotically the optimal fidelity.  As we noted in the multiphase covariant case,  the naive protocol consisting in measuring the optimal POVM and then re-preparing $M$ copies of the estimated state is strictly suboptimal, even in the asymptotic limit: indeed,    setting $K  = M$ one obtains   $$
F_{K= M }  [N\to M]  = \frac{F_{clon}  [N\to M]}{\sqrt 2} \, \left[ 1  +   O  \left(   \frac 1{\sqrt{ M^{1-\delta}}  }+    \frac {N^2} M    \right) \right]\, .
$$

\section{Cloning of two-qubit maximally entangled states}\label{sec:entangle}

In this section we consider the  $N$-to-$M$ cloning of a  two-qubit maximally entangled state, chosen at random according to the Haar measure.   In this case, the optimization of the cloning machine is much more challenging that it is for coherent,  phase-, and multiphase-covariant states,  even in the asymptotic limit of large $M$.  However, using the  bounding technique developed in the previous examples, we will be able to identify the optimal asymptotic cloner and to prove its global equivalence with state estimation.    Two bonus features of our optimal cloner are that \emph{i)} it is economical and \emph{ii)}  it can be implemented using only local operations on the entangled input systems, without resorting to classical communication or to global quantum operations.

\subsection{Decomposition of the input states}
Consider a general two-qubit maximally entangled state  $|\psi_g\>  \in  \spc H_A \otimes \spc H_B, ~  \spc H_A    \simeq  \spc H_B \simeq  \Cmplx^2$. We can imagine that the state $|\psi_g\>$ is shared by two parties, Alice and Bob, holding qubits $A$ and $B$, respectively.    The state can be parametrized as
\begin{equation*}
|\psi_g\>=\frac {(U_g\otimes I)|I\>\!\>}{\sqrt{2}}, \qquad g\in SU(2),
\end{equation*}
using the ``double-ket notation"  
$  |\Psi\>\!\>   :=   \sum_{m,n}    \< m  |  \Psi    |  n\>   ~  |m\>  |n\>$ for a generic operator $  \Psi$  \cite{DarianoLoPresti00}.    Using this notation, the state of  $N$ identical copies can be decomposed in a  convenient way.    First of all,  rearranging the Hilbert spaces in the tensor product, the input state  $|\psi_g\>^{\otimes N}$ can be considered as a vector in $ \spc H_A^{\otimes N}  \otimes  \spc H_B^{\otimes  N} $.      
Then, with a suitable choice of basis, the Hilbert spaces  $\spc H_A^{\otimes N}$ and   $\spc H_B^{\otimes N}$ can be decomposed as   direct sum of tensor product pairs as 
\begin{equation*}
\spc H_x^{\otimes N} = \bigoplus_{j   =j_{min}^{(N)}  }^{N/2}    \left(   \spc R^{(j,N)}_{x} \otimes \spc M^{(j,N)}_{x} \right) \qquad x  =  A, B 
\end{equation*} 
where    $j$ is the quantum number of the total angular momentum, $j_{min}^{(N)}=0$ for even $N$ and $j_{min}^{(N)}=\frac{1}{2}$ for odd $N$,   $ \spc R^{(j,N)}_{x}$ is a representation space, of dimension  $d_j = 2 j +1$, and $\spc M^{(j,N)}_{x}$ is a multiplicity space, of dimension  $$m_j^{(N)}=\frac{2d_j}{N+d_j+1}{N\choose N/2+j}$$    (see e.g. Ref. \cite{ChiribellaDariano05}). Relative to this decomposition, we can write $U_g^{\otimes N}$ as a block diagonal matrix with the blocks labelled by $j$, namely
\begin{equation}\label{cg}
U_g^{\otimes N}=\bigoplus_{j=j_{min}^{(N)}}^{N/2} \left[  U_{g}^{(j,N)}\otimes I_{m_j}^{(N)} \right].
\end{equation}
where $U_g^{(j,N)}   $ is the unitary operator representing the action of the element $g\in  SU(2)$ on the Hilbert space $\spc R^{(j,N)}_{x}$  and  $I_{m_j}^{(N)}$ denotes the identity on   $\spc M^{(j,N)}_{x}$.

Now, using  Eq. (\ref{cg}),   the input state   $|\psi_g\>^{\otimes N}$ can be cast in the form
\begin{eqnarray}
|\psi_g\>^{\otimes N}&=\frac  { |U^{\otimes N}_g\>\!\>}  {\sqrt{2^{N}}}   \\
&=\frac1  {\sqrt{  2^N}}\bigoplus_{j=j_{min}^{(N)}}^{N/2}   \left(|U_{g}^{(j,N)}\>\!\>\otimes|I_{m_j}^{(N)}\>\!\>\right)\\
\end{eqnarray}
with 
 $|U_{g}^{(j,N)}\>\!\>   \in \spc R^{(j,N)}_{A}  \otimes \spc R^{(j,N)}_{B}$ and  $|I_{m_j}^{(N)}\>\!\>   \in \spc M^{(j,N)}_{A} \otimes \spc M^{(j,N)}_{B} $.
 Hence, we obtained the decomposition 
\begin{equation}\label{decompositionmaxent} 
|\psi_g\>^{\otimes N}=\bigoplus_{j=j_{min}^{(N)}}^{N/2}\sqrt{p_{N,j}}  \, |\psi_{g}^{(j,N)}\>\\
\end{equation}
where
\begin{equation}\label{aaa}
|\psi_{g}^{(j,N)}\>:=\frac{|U_{g}^{(j,N)}\>\!\>}{\sqrt{d_j}}\otimes\frac{|I_{m_j}^{(N)}\>\!\>}{\sqrt{m_j^{(N)}}}
\end{equation}
and
\begin{eqnarray}
\label{ppp}   p_{N,j}  &:=\frac{d_j m_j^{(N)}}{2^N} =\frac{  2d_j^2}{N+d_j+1}  ~ B_{N,j} \, ,  
\end{eqnarray}
$B_{N,j}$ being the binomial distribution  $B_{N,j}  :=  {N \choose  N/2 + j}  /  2^N$.

\subsection{Upper bound on the cloning fidelity}
Following the strategy developed in the previous examples, we start our search  of the optimal asymptotic cloners by proving an upper bound on the global fidelity. The upper bound has the familiar form that appeared in propositions \ref{prop:uppercoh}, \ref{prop:upperphase} and \ref{prop:upperclock}, although its proof, provided in \ref{app:twoqubit}, requires a higher degree of  technicality:  
\begin{Proposition}\label{prop:twoqubit}   The global fidelity of the optimal   $N$-to-$M$ cloner of maximally entangled states is upper bounded as 
\begin{eqnarray}\label{fidbound}
F_{clon} [N \to M]  \le    \left\|   \rho_{AV}^{(M)}  \right\|_{\infty}  \,   p_{succ}^{(N)}       \, ,
\end{eqnarray}
where   $\left  \|    \rho^{(M)}_{AV} \right\|_{\infty}$ is the maximum eigenvalue of the average target state 
$  \rho_{AV}^{(M)}  :   =  \int   \d  g      \,   \psi_{g}^{\otimes M}  $ 
and $p_{succ}^{(N)}$ is the maximum probability density of correct identification of the input state, namely 
$   p_{succ}^{(N)}  :  = \max_{   \{  P_{g}\}}  \int    \d g \,       \< \psi_{g}|^{\otimes N}    P_{g}  \, |\psi_{g}\>^{\otimes N} \,$
the maximum running over all possible POVMs $\{  P_{g}\}$.
\end{Proposition}  
Similar to the proofs of the previous bounds, the proof here  is based on two ingredients:   the first is an upper bound on  the fidelity derived from the optimization over all possible covariant cloners, the second is a translation of the upper bound in terms of the quantities $\|  \rho_{AV}^{(M)}\|_\infty$ and $p_{succ}^{(N)}$.  Regarding $\|  \rho_{AV}^{(M)}\|_\infty$,  this can be easily computed using the Schur's lemma (cf. \ref{app:twoqubit}), which gives 
\begin{eqnarray}\label{avem}  \left \|   \rho_{AV}^{(M) }\right\|_{\infty}  = \frac{  p_{M ,  j^{(M)}_{\min}} }{d^2_{j_{\min}^{(M)}}}  \, .
\end{eqnarray} 
Regarding $p_{succ}^{(N)}$,  we recall the  expression of the optimal POVM  for the estimation of a maximally entangled state from $N$ copies, derived in  \cite{ChiribellaDariano05}, and given by
\begin{eqnarray}
\label{eta}   P_{{g}}:=|\eta_{\hat g}\>\<  \eta_{\hat g}| ,  \qquad  |\eta_{ g}\>    := U_{{g}}^{\otimes N}   |\eta\>   \qquad
|\eta\>:=\bigoplus_{j=j_{min}^{(N)}}^{N/2}  d_j   ~   |  \psi^{(j,N)}\> \, .  
\end{eqnarray}
Using this fact, we have
\begin{eqnarray}
\label{pns}     p^{(N)}_{succ}   & :  =     \<  \eta_g|      \psi_g^{\otimes N}  |\eta_g\>   =      \left (  \sum_{j=j_{min}^{(N)}}^{N/2}    \sqrt{  p_{N,j}}   \, d_j \right)^2  \qquad \forall g\in SU (2)\, .
\end{eqnarray}
The detailed derivation of Eq. (\ref{fidbound}) is provided in  \ref{app:twoqubit}.

\subsection{Asymptotically optimal economical cloner}
Here we exhibit an economical cloner that 
achieves the upper  bound of Eq. (\ref{fidbound}) in the limit of large $M$.   For simplicity, we will assume that $N$ and  $M$ are either both even or both odd.   In this case, our cloner consists simply in embedding the input state $  |\psi_g\>^{\otimes N}$ into the output space of $M$ copies.  This operation is described by  the isometry $V_{N,M}$ defined by the relation 
 \begin{eqnarray}\label{defining}
 V_{N,M}   \, |\psi_{g}^{(j,N)}\>   &  =  |\psi_{g}^{(j,M)}\>  \, ,
  \end{eqnarray}
 for every $g\in  SU(2)$ and for every $ j  \in  \{  j_{\min}^{(N)}, \dots,  N/2\}$.    It is worth mentioning that this economical cloner requires the use of global operations performed jointly on  Alice's and Bob's systems. Cloning of bipartite states under the restriction of Local Operations and Classical Communication (LOCC) has been recently considered by Kumagai and Hayashi \cite{kuma}, who investigated the asymptotic scenario where both $N$ and $M$ are large. The problem in Ref. \cite{kuma} was to copy a single, known bipartite state---a task that is made non-trivial by the LOCC restriction.  In our case, we allow general operations but require the cloning machine to work on arbitrary maximally entangled states.  
 
We now show that our cloner is asymptotically optimal.  Indeed,  its  fidelity  is given by 
\begin{equation}\label{ddd}
F [N \to M]=\left(\sum_{j=j_{min}^{(N)}}^{N/2}\sqrt{p_{N,j} \, p_{M,j}}\right)^2 \, .
\end{equation}
Now, we know  from Eq. (\ref{ppp})  that the ratio $p_{M,j}/d_j^2$ is proportional to the binomial  $B_{ M,j}$.  When $ M $ is large compared to $N$, the binomial is almost constant in the interval $[  j_{\min}^{(N)}, N/2]$ and  the fidelity can be approximated as   
\begin{eqnarray*}     F [N\to M]  &  =  \frac{ p_{M,  j_{\min}^{(M)}  } }{  d^2_{j_{\min}^{(M)}}}   \,  \left(\sum_{j=j_{min}^{(N)}}^{N/2}  \sqrt{ p_{N,j}  }   \, d_j \right)^2  \, \left[ 1  + O  \left(   \frac {N^2} M\right) \right]\\
  &  =   \left\|   \rho_{AV}^{(M)}  \right\|_{\infty}  \,   p_{succ}^{(N)}  \, \left[ 1  + O  \left(   \frac {N^2} M\right) \right]     
\end{eqnarray*}
having used Eqs. (\ref{avem}) and (\ref{pns}).  
This means that, asymptotically,  our economical cloner saturates the upper bound of Eq. (\ref{fidbound}).  

\medskip 
\noindent {\bf  Remark  (large $N$ asymptotics). }   For large $N$, the Gaussian approximation for the binomial allows one to approximate the probability distribution $p_{N,j}$ in  Eq. (\ref{ppp}) as 
\begin{eqnarray}\label{approx}
 p_{N,j}  =  \sqrt{\frac{2}  {\pi  N^3} }    \,  2(2j+1)^2  \exp\left[  - \frac{  2  j^2} N\right] \left[ 1+  O\left(  \frac 1  {\sqrt{N^{1-\delta}}}\right) \right] + \eta_N  \,   
\end{eqnarray} 
for arbitrary  $\delta>0$ and $\eta_N$ vanishing faster than the inverse of any polynomial. Approximating the summation in Eq. (\ref{ddd}) with a Gaussian integral,  one can  find the close form expression 
\begin{eqnarray*}
F [N \to M] &=  
\left(\frac{  \sqrt{ 4MN}}{M+N}\right)^3  \left[ 1+  O\left(  \frac 1  {\sqrt{N^{1-\delta}}}\right) \right] .
\end{eqnarray*} 
Intriguingly, the above fidelity is equal to the fidelity of the economical cloner for multiphase covariant states in dimension $d=4$.  In both cases, the fidelity of the economical  cloner has the form 
$$ F [N \to M]  =  \left(\frac{  \sqrt{ 4MN}}{M+N}\right)^f \, \left[ 1+  O\left(  \frac 1  {\sqrt{N^{1-\delta}}}\right) \right]  , $$
where $f$ is the number of free parameters needed to specify the input state.   This observation, along with the observations made for phase- and multiphase- covariant cloning,  suggests that the optimal economical cloners satisfy a universality property, which forces their fidelity to depend only on the numbers of free parameters, and not on the specific details of the states to be cloned.


\subsection{Asymptotically optimal MP protocol}
Now we are going to show how to reach the optimal cloning fidelity  with a suitable MP  protocol.  By now, the choice of the protocol should be obvious:  estimate the state using the optimal POVM of Eq. (\ref{eta}), prepare $K  =  \lceil  M^{1-\epsilon}\rceil$ copies according to the estimate, and clone from $K$ to $M$ using the cloner of Eq. (\ref{defining}).   The calculation of the fidelity, done in  \ref{app:MP2qubit}, gives the value
\begin{eqnarray}\label{KK}
 F_K [  N \to M]  &  =   p_{succ}^{(N)}    \,     \sqrt{  \frac{  8} {  \pi  (  M+K)^3}}  \,  \left[ 1+  O\left(  \frac 1  {\sqrt{K^{1-\delta}}}  +  \frac NK\right) \right] .    
\end{eqnarray}
Now, choosing $K  =  \lceil  M^{1-\epsilon}  \rceil$, in the large $M$ limit we have 
\begin{eqnarray*}
     \sqrt{  \frac{  8} {  \pi  (  M+K)^3}}      &  =   \frac{ 2 B_{  M,0}}{M}  \left[   1  +   O\left(  \frac K  M \right) \right]  \\
   &  =   \left \| \rho_{AV}^{(M)} \right\|_{\infty} \,   \left[   1  +   O\left(  \frac K  M \right) \right]   \, ,
\end{eqnarray*}
and, therefore 
\begin{eqnarray*}
F_{  K}  [  N \to M]  &=      p_{succ}^{(N)}    \,    \left \| \rho_{AV}^{(M)} \right\|_{\infty}  \, \left[ 1+O\left( \frac 1  {\sqrt{K^{1-\delta}}}  + \frac N K +  \frac K M\right) \right]  .
\end{eqnarray*}

This relation shows that asymptotically,  the fidelity of our protocol is arbitrarily close to the fidelity of the optimal cloner.        Once more, note that re-preparing   $M$ identical copies is strictly suboptimal, even in the asymptotic limit.   Indeed, setting $K  = M$ in Eq. (\ref{KK}) we obtain 
$$  F_M  [  N \to M]  = \frac{F_{clon}[N\to M]}{\sqrt {2^3}} \, \left[ 1+O\left( \frac 1  {\sqrt{M^{1-\delta}}}  +  \frac {N^2} M \right) \right]  .$$   
Note that the ratio between the fidelity of the naive MP protocol and that of the optimal cloner has the same value of the ratio in the case of multiphase-covariant cloners in dimension $d=4$.  Again, it appears that  the optimal economical cloners enjoy a universality property, where the relevant quantities depend only on the number of free parameters needed to describe the input state. 


\section{Impossibility of a global equivalence in terms of channel distance}\label{sec:distance}
In the previous sections we showed a wealth of examples where  the optimal asymptotic cloners are economical.  
Before concluding,  we show that every economical channel  can be distinguished well from every MP channel, in the sense that the probability of error is upper bounded by a finite value bounded away from 1/2.   This result leads to an important \emph{caveat}:   the equivalence between cloning and estimation, valid at the level of fidelities, cannot be valid at the level of channels, because there exist optimal cloning channels that never come close to the set of MP channels. This situation contrasts sharply with the single-copy scenario, where the distance between the single-copy restrictions of the optimal cloner converges to the single-copy restrictions of an MP channel  \cite{Chiribella11}.

\subsection{On the distance between economical channels and MP channels}
Here we give a precise quantitative meaning to the the statement that  economical channels are far from MP channels.   As a distance measure for channels, we use the trace distance, which for two channels $\map C$ and $\widetilde{\map C}$ is defined as  
\begin{eqnarray}
\|  \map C  -  \widetilde{\map C}  \|_1  : =   \max_{|\psi\> \in\spc H_{in}, \| | \psi\> \| =1  }    \left\|   \map C (\psi)   - \widetilde{ \map   C}   (\psi) \right \|_1,   
\end{eqnarray}  
where  $\spc H_{in}$ is the input Hilbert space of channels $\map C$ and $\widetilde{\map C}$ and $\|  A  \| _1 :=  \Tr   |A| $ denotes  the trace norm of the operator $A$.   The operational meaning of the trace distance between two channels is  provided by Helstrom's theorem on minimum error discrimination:  if one tries to discriminate between the two channels $\map C$ and $\widetilde{\map C}$, given with prior probabilities $ p$ and $1-p$, respectively, one can achieve the average probability of error  
$$p_{err}  =  \frac{  1  -  \|  p\,  \map C- (1-p)  \widetilde{\map C}\|_1  }2\, ,$$  by choosing the best input state in $\spc H_{in}$.      This means that   when the trace distance is close to 1,  the two channels $\map C$ and $\widetilde{\map C}$ are almost perfectly distinguishable, even without using the assistance of additional ancillas.   A simple bound on the trace-distance between an economical channel and an MP channel is given by the following:  
\begin{Proposition}\label{distance}  Let $ \map C$ and $\widetilde {\map C}$ be an economical channel and an MP channel, transforming density matrices on $\spc H_{in}$ into density matrices on $\spc H_{out}$, respectively.     The trace distance between $\map C$
 and $\widetilde {\map C}$ is lower bounded as 
 \begin{eqnarray}\label{distancebound}
 \|  \map C  - \widetilde { \map C} \|_1  \ge   2  \left(1  -	d^{-1}_{in}\right).   
 \end{eqnarray}
 \end{Proposition}
The proof is  provided  in \ref{app:distance}.  Note that the lower bound   tends to the maximum possible value $  \|  \map C  -  \widetilde {\map C} \|_1  =2$  when $d_{in}$ is large, meaning that for large input spaces economical channels and MP protocols  produce almost orthogonal output states. Due to the above bound,  there is no way to approximate an economical cloning channel with an MP protocol, even in the  macroscopic limit $M\to \infty$.  
 Hence,   proving the global asymptotic equivalence with state estimation means proving a non-trivial statement about the performances of two radically different types of processes.

\section{Discussion and conclusions}\label{sec:conclusion}

In this paper we investigated the asymptotic behaviour of quantum copy machines. We posed  the question whether  asymptotic cloning is equivalent to state estimation  when all the clones are considered jointly. 
We conjectured that the global equivalence holds and we proved its validity for arbitrary finite sets of states,  arbitrary families of coherent states,  arbitrary phase- and multiphase-covariant sets of states, and for two qubit maximally entangled states.   These examples indicate that,  if a counterexample to our conjecture exists at all, it should involve a rather exotic set of states.  

Interestingly, in some of our examples the optimal asymptotic performance can be achieved by an economical cloner, which requires a single unitary interaction and no extra ancillas (except for the $M-N$ blank copies on which the information is redistributed).  Here the fact that both economical cloners and MP protocols are asymptotically optimal establishes a non-trivial equality between the performances of two very different ways to process information.   The examples where economical cloners are asymptotically optimal include cases where the input states are generated by a commuting set of unitaries, like phase- and multiphase-covariant cloning, as well as cases where the set is not-commuting, like the cloning of two-qubit maximally entangled states. It is then natural to ask which properties of the set of input states are responsible for the asymptotic optimality of economical cloners.

Two remarkable features emerged from our analysis, along with new intriguing questions.    First, whenever the set of states is generated by the action of a group and the prior probability is uniform, we found that the cloning fidelity is upper bounded by a simple function of the likelihood, i.e. the probability density of successful identification of the input state.  Precisely, one has 
\begin{eqnarray}\label{clonebound}
F_{clon} [N \to M]  \le    \left\|   \rho_{AV}^{(M)}  \right\|_{\infty}  \,   p_{succ}^{(N)}       \, ,
\end{eqnarray}
where $p_{succ}^{(N)}$ is the likelihood and    $\|   \rho_{AV}^{(M)}  \|_{\infty}$ is the maximum eigenvalue of the average target state.   It is then natural to ask whether Eq. (\ref{clonebound}) holds for every cloning problem in the presence of symmetry.     
There are several reasons why this would be desirable.    The first reason is that a bound on the cloning fidelity in terms of the likelihood of estimation  is useful \emph{per se}, independently of the question about the asymptotic equivalence.  Indeed, this bound establishes a bridge between quantum cloning and maximum likelihood measurements, which have been studied extensively in the literature  \cite{Helstrom,Holevo,ChiribellaDariano04,ChiribellaDarianoPerinotti06}. Since these works  provide a closed expression for $p_{succ}^{(N)}$, Eq. (\ref{clonebound})  would  be an easily computable upper bound on the cloning fidelity, which could be readily used for the design of nearly-optimal cloning machines.  In fact, we expect that Eq. (\ref{clonebound}) will hold not only for cloning, but also for a variety of other transformations of resources in the presence of symmetry \cite{gour2008,marvian2013,skotiniotis2012,ahmadi2013,varun}.


Another remarkable feature emerging from our work is that the fidelities of the optimal asymptotic cloners obey a universality property: when both $N$ and $M$ are large, the fidelity becomes independent of the specific details of input states and scales as
\begin{eqnarray}\label{universal}
F_{clon}  [N\to M]  =   O  \left[  \left(\frac NM\right)^{f/2}  \right]
\end{eqnarray}    
 for all families of states described by $f$ free real parameters. 
This fact was observed explicitly for phase- and multiphase-covariant cloning and for the cloning of two-qubit maximally entangled states, but  can be easily seen to hold also for several cases of general coherent states: For example, for the harmonic oscillator coherent states $\{ |\alpha\>\}_{\alpha  \in  \mathbb C}$  the (worst-case) cloning fidelity is $ F^*_{clon}  [N \to M]  = N/M$  \cite{CerfIpe00,CerfIblisdir00},  consistently with Eq. (\ref{universal}) and with the fact that the state $|\alpha\>$ is described by two real parameters.  Similarly,  for the pure states in dimension $d$ the optimal cloning fidelity  \cite{Werner98} satisfies
\begin{eqnarray*} F_{clon}  [N\to M]   
 & =   \left(\frac N  M  \right)^{d-1}    \left[ 1+   O   \left(  \frac 1 N \right)\right]  \, ,
 \end{eqnarray*}  
consistently with Eq. (\ref{universal}) and with the fact that an arbitrary pure state is described by $2(d-1)$ free parameters.  The universality property expressed by Eq. (\ref{universal}) appears as a deep fact about asymptotic cloning machines, and it is our hope that our work will stimulate research in this direction, eventually leading to a general proof.

\subparagraph*{Acknowledgements} This work was supported in part by the National Basic Research Program of China through Grant 2011CBA00300, 2011CBA00301, by the National Natural Science Foundation of China through Grants 11350110207, 61033001, 61361136003, by FQXi through the large grant ``The fundamental principles of information dynamics", and by the 1000 Youth Fellowship Program of China.  GC acknowledges  discussions with E Bagan about the limits of applicability of the central limit theorem.

\medskip 

\begin{thebibliography}{10}
\expandafter\ifx\csname url\endcsname\relax
  \def\url#1{{\tt #1}}\fi
\expandafter\ifx\csname urlprefix\endcsname\relax\def\urlprefix{URL }\fi
\providecommand{\eprint}[2][]{\url{#2}}

\bibitem{ScaraniIblisdir05}
Scarani V, Iblisdir S, Gisin N and Ac{\'\i}n A 2005 {\em Rev. Mod. Phys.\/}
  {\bf 77} 1225--1256

\bibitem{CerfFiurasek06}
Cerf N~J and Fiur{\'a}{\v{s}}ek J 2006 {\em Prog. Optics\/} {\bf 49} 455--545

\bibitem{Helstrom}
Helstrom C~W 1976 {\em Quantum detection and estimation theory\/} vol~84 (New
  York: Academic press)

\bibitem{Holevo}
Holevo A~S 2011 {\em Probabilistic and statistical aspects of quantum theory\/}
  vol~1 (Basel: Springer-Verlag)

\bibitem{Werner01}
Werner R~F 2001 Quantum information theory: an invitation {\em Quantum
  Information\/} ({\em Springer Tracts in Modern Physics\/} vol 173) (Berlin
  Heidelberg: Springer-Verlag) pp 14--57 ISBN 978-3-540-41666-1

\bibitem{Keyl02}
Keyl M 2002 {\em Phys. Rep.\/} {\bf 369}(5) 431--548

\bibitem{Hayashi}
Hayashi M 2007 {\em Quantum Information: An Introduction\/} (Berlin Heidelberg:
  Springer-Verlag)

\bibitem{GisinMassar97}
Gisin N and Massar S 1997 {\em Phys. Rev. Lett.\/} {\bf 79}(11) 2153--2156

\bibitem{BrussEkert98}
Bru\ss{} D, Ekert A and Macchiavello C 1998 {\em Phys. Rev. Lett.\/} {\bf
  81}(12) 2598--2601

\bibitem{Werner98}
Werner R~F 1998 {\em Phys. Rev. A\/} {\bf 58} 1827--1832

\bibitem{BrussDivincenzo98}
Bru\ss{} D, DiVincenzo D~P, Ekert A, Fuchs C~A, Macchiavello C and Smolin J~A
  1998 {\em Phys. Rev. A\/} {\bf 57} 2368--2378

\bibitem{CerfIpe00}
Cerf N~J, Ipe A and Rottenberg X 2000 {\em Phys. Rev. Lett.\/} {\bf 85}
  1754--1757

\bibitem{CerfIblisdir00}
Cerf N~J and Iblisdir S 2000 {\em Phys. Rev. A\/} {\bf 62}(4) 040301

\bibitem{namiki1}
Namiki R 2011 {\em Phys. Rev. A\/} {\bf 83} 040302(R)

\bibitem{namiki2}
Namiki R, Koashi M and Imoto N 2008 {\em Phys. Rev. Lett.\/} {\bf 101} 100502

\bibitem{ChiribellaXie13}
Chiribella G and Xie J 2013 {\em Phys. Rev. Lett.\/} {\bf 110} 213602

\bibitem{KeylWebsite}
Keyl M Asymptotic cloning is state estimation
  \url{http://qig.itp.uni-hannover.de/qiproblems/22}

\bibitem{BaeAcin06}
Bae J and Ac{\'\i}n A 2006 {\em Phys. Rev. Lett.\/} {\bf 97}(3) 030402

\bibitem{ChiribellaDariano06}
Chiribella G and D'Ariano G~M 2006 {\em Phys. Rev. Lett.\/} {\bf 97}(25) 250503

\bibitem{Chiribella11}
Chiribella G 2011 {\em Proc. TQC 2010, Lecture Notes in Computer Science\/}
  {\bf 6519/2011} 9--25

\bibitem{Wiesner83}
Wiesner S 1983 {\em ACM Sigact News\/} {\bf 15}(1) 78--88

\bibitem{Aaronson09}
Aaronson S 2009 {\em 24th Annual IEEE Conference on Computational Complexity\/}
   229--242

\bibitem{FarhiGosset12}
Farhi E, Gosset D, Hassidim A, Lutomirski A and Shor P 2012 {\em Proceedings of
  the 3rd Innovations in Theoretical Computer Science Conference\/}  276--289

\bibitem{MolinaVidick13}
Molina A, Vidick T and Watrous J 2013 {\em Proceedings of the Conference on
  Theory of Quantum Computation, Communication, and Cryptography\/}  45--64

\bibitem{HammererWolf05}
Hammerer K, Wolf M~M, Polzik E~S and Cirac J~I 2005 {\em Phys. Rev. Lett.\/}
  {\bf 94}(15) 150503

\bibitem{AdessoChiribella08}
Adesso G and Chiribella G 2008 {\em Phys. Rev. Lett.\/} {\bf 100}(17) 170503

\bibitem{OwariPlenio08}
Owari M, Plenio M~B, Polzik E~S, Serafini A and Wolf M~M 2008 {\em New J.
  Phys.\/} {\bf 10}(11) 113014

\bibitem{CalsamigliaAspachs09}
Calsamiglia J, Aspachs M, Mu\~noz-Tapia R and Bagan E 2009 {\em Phys. Rev. A\/}
  {\bf 79} 050301

\bibitem{Namiki11}
Namiki R, Koashi M and Imoto N 2008 {\em Phys. Rev. Lett.\/} {\bf 101} 100502

\bibitem{ChiribellaAdesso14}
Chiribella G and Adesso G 2014 {\em Phys. Rev. Lett.\/} {\bf 112} 010501

\bibitem{DemartiniSciarrino08}
De Martini F, Sciarrino F and Vitelli C 2008 {\em Phys. Rev. Lett.\/} {\bf 100}
  253601

\bibitem{DemartiniSciarrino12}
De~Martini F and Sciarrino F 2012 {\em Rev. Mod. Phys.\/} {\bf 84}(4)
  1765--1789

\bibitem{NiuGriffiths99}
Niu C~S and Griffiths R~B 1999 {\em Phys. Rev. A\/} {\bf 60}(4) 2764

\bibitem{ChiribellaYang13}
Chiribella G, Yang Y and Yao A~C~C 2013 {\em Nat. Comm.\/} {\bf 4}(2915)

\bibitem{Gilmore72}
Gilmore R 1972 {\em Ann. Phys.\/} {\bf 74}(2) 391--463

\bibitem{Perelomov}
Perelomov A 1986 {\em Generalized coherent states and their applications\/}
  (New York: Springer-Verlag)

\bibitem{Perelomov72}
Perelomov A 1972 {\em Comm. Math. Phys.\/} {\bf 26}(3) 222--236

\bibitem{aliAntoine}
Ali S~T, Antoine J~P and Gazeau J~P 2000 {\em Coherent states, wavelets and
  their generalizations\/} (New York: Springer-Verlag)

\bibitem{ChiribellaDarianoLorentz06}
Chiribella G, D’Ariano G~M and Perinotti P 2006 {\em Laser Phys.\/} {\bf
  16}(11) 1572--1581

\bibitem{ChiribellaDariano04}
Chiribella G, D’Ariano G~M, Perinotti P and Sacchi M~F 2004 {\em Phys. Rev.
  A\/} {\bf 70}(6) 062105

\bibitem{ChiribellaDarianoPerinotti06}
Chiribella G, D'Ariano G~M, Perinotti P and Sacchi M~F 2006 {\em Int. J. Quant.
  Info.\/} {\bf 4}(03) 453--472

\bibitem{ChiribellaAdesso13}
Chiribella G and Adesso G 2014 {\em Phys. Rev. Lett.\/} {\bf 112}(1) 010501

\bibitem{BrussCinchetti00}
Bru\ss{} D, Cinchetti M, D'Ariano G~M and Macchiavello C 2000 {\em Phys. Rev.
  A\/} {\bf 62} 012302

\bibitem{FanMatsumoto01}
Fan H, Matsumoto K, Wang X~B and Wadati M 2001 {\em Phys. Rev. A\/} {\bf 65}
  012304

\bibitem{FanMatsumoto02}
Fan H, Matsumoto K, Wang X~B and Imai H 2002 {\em J. Phys. A\/} {\bf 35}(34)
  7415

\bibitem{DarianoMacchiavello03}
D'Ariano G~M and Macchiavello C 2003 {\em Phys. Rev. A\/} {\bf 67}(4) 042306

\bibitem{BuscemiDariano05}
Buscemi F, D'Ariano G~M and Macchiavello C 2005 {\em Phys. Rev. A\/} {\bf
  71}(4) 042327

\bibitem{DurtFiurasek05}
Durt T, Fiur{\'a}{\v{s}}ek J and Cerf N~J 2005 {\em Phys. Rev. A\/} {\bf 72}(5)
  052322

\bibitem{DarianoLoPresti00}
D'Ariano G~M, Presti P~L and Sacchi M~F 2000 {\em Phys. Lett. A\/} {\bf 272}(1)
  32--38

\bibitem{ChiribellaDariano05}
Chiribella G, D'Ariano G~M and Sacchi M~F 2005 {\em Phys. Rev. A\/} {\bf 72}
  042338

\bibitem{kuma}
Kumagai W and Hayashi M 2013 {\em arXiv:1306.4166\/}

\bibitem{gour2008}
Gour G and Spekkens R~W 2008 {\em New J. Phys.\/} {\bf 10}(3) 033023

\bibitem{marvian2013}
Marvian I and Spekkens R~W 2013 {\em New. J. Phys.\/} {\bf 15}(3) 033001

\bibitem{skotiniotis2012}
Skotiniotis M and Gour G 2012 {\em New. J. Phys.\/} {\bf 14}(7) 073022

\bibitem{ahmadi2013}
Ahmadi M, Jennings D and Rudolph T 2013 {\em New J. Phys.\/} {\bf 15}(1) 013057

\bibitem{varun}
Narasimhachar V and Gour G 2014 {\em Phys. Rev. A\/} {\bf 89}(3) 033859

\end{thebibliography}
\providecommand{\newblock}{}

\appendix
\section{Approximating a finite set of quantum states with an orthonormal set}\label{app:gram}

Here we consider the Gram-Schmidt orthogonalization procedure, which transforms a set of linearly independent  unit vectors $\{  |\psi_x\>\}_{x\in\mathsf X}$ into a set of orthonormal vectors $\{  |\gamma_x\>\}_{x\in\mathsf X}$, and we ask  how far is $|\gamma_x\>$  from the original vector $|\psi_x\>  $.  An upper bound on the distance is provided by the following
\begin{Theo}
For an arbitrary finite set of unit vectors $\{ |\psi_x\>\}_{x  \in \mathsf X}$ there is a set of orthonormal states $\{  |\gamma_x\>\}_{x\in \mathsf X}$ such that, for every $x\in\mathsf X$,
\begin{eqnarray*}
\left \|    \psi_x  -  \gamma_x  \right \|_\infty  \le   \sqrt{ \frac{\eta \, \alpha^ {|\mathsf X|}  }{\alpha-1}  }  \, , 
\end{eqnarray*}
where $\eta  :=  \max_{x\not = y} |\< \psi_x|\psi_y\>  |^2 $ and $ \alpha  =  3  + 2 \sqrt{2} $. 
\end{Theo}  

{\bf Proof.}  By definition, one has  $ \|   \psi_x  -  \gamma_x  \|_{\infty}  =   \sqrt{  1  -  |\<  \psi_x|\gamma_x\>|^2}  $.
 Hence, to prove the thesis we only need to find a set of orthonormal vectors such that, for every $n$, the moduli of the scalar products are lower bounded as 
 \begin{eqnarray}\label{infnorm} 
 |\<  \psi_x|\gamma_x\>|^2      \ge 1  - \frac{\eta \, \alpha^ {|\mathsf X|}  }{\alpha-1}  \qquad \forall x\in\mathsf X  \, .
 \end{eqnarray}   
 This is accomplished by the Gram-Schmidt orthonormalization procedure,  defining 
\begin{eqnarray*}
|\gamma_1\> &:  =  |\psi_1\>  \\
|\gamma_{x+1}\>  &  :  =   \frac{  |\psi_{x+1}\>  -   \sum_{y=1}^x    \<  \gamma_y | \psi_{x+1}\>  ~  |\gamma_y\>   }{\sqrt{  1  -  \sum_{y=1}^x    |\<  \gamma_y| \psi_{x+1}\>  |^2 }}  \, .    
\end{eqnarray*}
We now lower bound the scalar product between a state and the corresponding Gram-Schmidt vector  as 
\begin{eqnarray}\label{infnorm2}
| \<  \psi_x  |  \gamma_x\> |^2  \ge  1-  c_x \eta \, ,
\end{eqnarray} 
and prove the upper bound  
$$c_x  \le \frac{ \alpha^ {|\mathsf X|}  }{\alpha-1}  \qquad \forall x \in\mathsf X  \, .$$ 
We proceed by induction, starting from the observation that, by construction, we can set $c_1 = 0$.        To continue the induction, note that  
\begin{eqnarray*}  
\<  \psi_{x+1}  |  \gamma_{x+1}\>    & =   \sqrt{  1  -   \sum_{y=1}^x    |\<  \gamma_y| \psi_{x+1}\>  |^2 } ,
\end{eqnarray*} 
and, for every $y  \le x$,  
\begin{eqnarray*}
|\<  \gamma_y| \psi_{x+1}\>  |  
&\le    \left   | \<  \gamma_y| \psi_{x+1}\>  -  \<  \psi_y| \psi_{x+1}\>   \right|  +   |\<  \psi_y| \psi_{x+1}\>  | \\
&\le    \left\|    | \gamma_y\>  -  |\psi_y\>  \right\|  +  \sqrt{\eta}   \\
&=     \sqrt{2  -   2\<  \gamma_y  | \psi_x\>   } +   \sqrt{\eta}   \\
&\le    \sqrt{ 2   - 2  \<  \gamma_y  | \psi_y\>^2  }  + \sqrt\eta  \\
&\le    \left( \sqrt{2c_y}  +  1  \right)   \sqrt{\eta}  \, .
\end{eqnarray*}
Hence, the scalar product between $|\psi_{x+1}\>$ and $  |\gamma_{x+1}\>$ is lower bounded as
\begin{eqnarray*}
|\<  \gamma_{x+1}| \psi_{x+1}\>  |^2  &\ge 1  -  \sum_{y=1}^{x}  \left(\sqrt{2c_y}+1\right)^2 \eta \, ,
\end{eqnarray*}
which means that we can choose  $c_{x+1}  =  \sum_{y=1}^x    \left(\sqrt{2c_y}+1\right)^2$. This choice  gives the recursion relation $c_{x+1}  =  c_x  +     \left(\sqrt{2c_x}+1\right)^2 $ and the bound
\begin{eqnarray}\label{recursion}  c_{x+1}  \le \alpha c_x +  1    \qquad  \alpha: = 3  +   2\sqrt{2} \, , 
\end{eqnarray}
valid both for $x=1$   (where $c_1= 0)$ and for  $x > 1$ (where $c_x  \ge 1$).  Now, we have $c_1  = 0$,   
$c_2 = 1$, and, for for $x  \ge 2$, Eq. (\ref{recursion}) implies 
 \begin{eqnarray*}
c_{x} &   \le  \sum_{y=0}^{x-2} \alpha^{y}   = \frac{ \alpha^{x-1}-1}{\alpha  -1}   \le \frac{ \alpha^{x-1}}{\alpha  -1}  \le  \frac{ \alpha^{|\mathsf X|}}{\alpha  -1}
\end{eqnarray*}  
Combining this bound with  Eq. (\ref{infnorm2}) then obtain   Eq. (\ref{infnorm}), concluding the proof.  
 \qed 

\section{Upper bound on the cloning fidelity for multiphase covariant states}\label{app:proofuppermultiphase}
 {\bf Proof of proposition \ref{prop:upperphase}.}  Let $\map C_{N,M}$ be the optimal channel.  Writing it in the Kraus form and imposing the covariance condition, it is simple to verify that  the action of the channel on the input state $|\psi\>^{\otimes N}$ is 
\begin{eqnarray*}
\map C_{N,M}  \left( \psi^{\otimes N}\right)      = \sum_{\vec \mu}   ~ &  \sum_{ \vec n, \vec n' \in  \map P_{N,d}}     \sqrt{p_{N,\vec n}  p_{N,\vec n'}}  ~    c^{\vec \mu}_{\vec n  \vec n'}  \\
&    \times |M,\vec n  + \vec  \mu\>\<  M, \vec n'+ \vec  \mu|  \,  ,
\end{eqnarray*}
   where $\vec \mu$ is a vector of integers satisfying $  \sum_{j=0}^{d-1}   \mu_j  =  M-N$ and   $  c^{\vec \mu}_{\vec n\vec n'}$   is a positive matrix satisfying the normalization condition 
   \begin{eqnarray}\label{subject}\sum_{\vec\mu}   c^{\vec\mu}_{\vec n\vec n}  = 1\qquad \forall \vec n  \in \map P_{N,d} \,. 
   \end{eqnarray} 
Using this fact, we can upper bound the fidelity as 
\begin{eqnarray}
\nonumber F_{clon}[N \to M]  &  =   \< \psi|^{\otimes M} \, \map C_{N,M}  \left( \psi^{\otimes N}\right)   \,  |  \psi\>^{\otimes M} \\
\nonumber    &  =   \sum_{\vec \mu} \sum_{ \vec n, \vec n' }    \sqrt{  p_{M,\vec n+\vec \mu} \,  p_{M,\vec n'+ \vec  \mu}  \, p_{N,\vec n} \,  p_{N,\vec n'}}\,    c^{\vec \mu}_{\vec n\vec  n'}  \\
\nonumber &  \le   \sum_{\mu}    \left( \sum_{ \vec n }     \sqrt{   p_{M,\vec n+\vec \mu}  \, p_{N,\vec n} \, c^{\vec \mu}_{\vec n\vec n}}   \right)^2 \\
\nonumber &  \le        p_{M, \vec m_*}     \sum_{\mu}   \left( \sum_{\vec  n } \sqrt{p_{N,\vec n} \,   \, c^{\vec \mu}_{\vec n\vec n}}   \right)^2  \, ,
\end{eqnarray}
where $\vec m_*$ is the partition that maximizes $ p_{M,\vec m}$.  
 We can now maximize the r.h.s. of the bound over the coefficients $c^{\vec\mu}_{{\vec n}{\vec n}}$, subject to the constraint of Eq.  (\ref{subject}).    Using the method of Lagrange multipliers, it is immediate to obtain the bound
\begin{eqnarray*}
F[N\to M]    \le        \,   p_{  M,  \vec m_*}      \,   \left(  \sum_{\vec n} \sqrt{  p_{N,\vec n}   }  \right)^2 \, . 
\end{eqnarray*}
Using Eqs.  (\ref{rhoavmax}) and (\ref{psuccmultiphase}) it is  immediate to recognize in the r.h.s. the bound promised in Eq. (\ref{upperphase}).   \qed

\section{Upper bound on the cloning fidelity for two-qubit maximally entangled states}\label{app:twoqubit}

{\bf Proof of proposition \ref{prop:twoqubit}.}     
Using the notation $  \spc H  \otimes |  \beta\>$ to denote the subspace spanned by vectors of the form $  |\alpha\>  |\beta\>$, with $  |\alpha\>  \in  \spc H$, we have that  every state $ |  \psi_g\>^{\otimes N}$ in Eq. (\ref{aaa}) belongs to the  subspace 
\begin{eqnarray}
\nonumber \spc H^{(N)}_{ent}    &:  =   \bigoplus_{j={j_{min}^{(N)}}}^{N/2}  \left( \spc R_A^{(j,N)}  \otimes  \spc R_B^{(j,N)}   \otimes    |  I_{m_j}^{(N)}  \>\!\> \right) \\
 &  \subset \bigoplus_{j={j_{min}^{(N)}}}^{N/2}  \left( \spc R_A^{(j,N)}  \otimes  \spc R_B^{(j,N)}   \otimes  \spc M_A^{(j,N)}  \otimes  \spc M_B^{(j,N)}\right) 
\label{spacedecomposition}
\end{eqnarray}   
Hence, for the optimization of the fidelity  we can restrict our attention to this subspace and consider quantum channels that map  states on $\spc H^{(N)}_{ent}$ to states on $\spc H^{(M)}_{ent}$.  As usual in the presence of symmetry, the optimization can be restricted without loss of generality to the set of covariant cloners, which satisfy the condition    $$   \map C_{N,M}    \left( \map U_g  \otimes \map U_h \right)^{\otimes N}   =     \left( \map U_g  \otimes \map U_h \right)^{\otimes M}    \map C_{N,M}    \qquad \forall g,h \in  SU(2)\, . $$     
For simplicity, we focus on the case where $N$ and $M$ have the same parity (note that this restriction does not make any difference in the asymptotic limit). In order to find the optimal channel, it is useful to translate the problem in terms of Choi operators. In this language, the channel is represented by a positive operator $C_{N,M}$  acting on $  \spc H_{ent}^{(M)}  \otimes \spc H_{ent}^{(N)}$ and the covariance condition becomes the commutation relation   
\begin{eqnarray}\label{comm} \left  [ C_{N,M}   ,     \left(U_g  \otimes U_h \right)^{\otimes M}  \otimes \left(U^*_g  \otimes U^*_h \right)^{\otimes N}      \right]  =  0 \, ,   
\end{eqnarray}  
where $U^*$ denotes the entry-wise complex conjugate of the matrix $U$.  The fidelity reads 
$$  F[N  \to M]  =    \<  \psi|^{\otimes (M+N)}   C_{N, M}  |  \psi\>^{\otimes (M+N)}  $$      
and has to be maximized under the constraint  of trace-preservation 
\begin{eqnarray}\label{ttt} \Tr_{M}  [  C_{N,M}]  =  I_N \, ,
\end{eqnarray}
where $\Tr_M$   ($I_N$) denotes the trace over (identity operator on) the output (input) Hilbert space. 

Now, Eq. (\ref{aaa}) allows us to express the state $  |\psi\>^{\otimes ( M+  N)}$ as 
\begin{eqnarray}\label{bbb}    & |\psi\>^{\otimes (M+N)}   =     \bigoplus_{l= l^{(  M+N)}_{\min}}^{ ( M+  N)/2}  |I_l\>\!\>   |  \alpha_l\>   \\
 \nonumber  &|\alpha_l\>    :=  \bigoplus_{(j,k) \to l }   \sqrt{ \frac {  p_{M,j}  p_{  N,  k}   }{  d_j d_k } }  ~  \frac{  |  I^{(M)}_{m_j}\>\!\> }{ \sqrt{  m^{(M)}_j} }  \otimes \frac{  |  I^{(N)}_{m_k}\>\!\> }{ \sqrt{  m^{(N)}_k} }  \, ,   
\end{eqnarray}
where the notation $  (  j,k)  \to l$ denotes the pairs $(j,k)$ that add up to $l$ by the addition rules of the angular momenta.  Combining the above expression with the commutation relation of Eq. (\ref{comm}), it is easy to prove that, without loss of generality, the optimal Choi operator can be chosen to be of the form  
\begin{eqnarray}
\nonumber C_{N,M}     &=  \bigoplus_{l}   \left(   I_l  \otimes I_l  \otimes A_l\right)\\ 
 A_l    &=  \sum_{\begin{array}{ll} (j,k)&\to l\\
  (j'l') &\to k
  \end{array}}   [A_{l}]_{(j,k)  (j',k')}     
\,  \frac{  |  I^{(M)}_{m_j}\>\!\> \<\!\< I^{(M)}_{m_{j'}}| }{  \sqrt{ m^{(M)}_j  m^{(M)}_{j'} }}  \otimes \frac{  |  I^{(N)}_{m_k}\>\!\>\<\!\< I^{(N)}_{m_{k'}}|  }{  \sqrt{m^{(N)}_k  m^{(N)}_{k'}   } } \, , 
\label{ccc}   
\end{eqnarray}
where   $   [A_{l}]_{(j,k)  (j',k')}  $ is a positive matrix.  
Combining Eqs. (\ref{bbb}) and (\ref{ccc}), the fidelity can be upper bounded as 
\begin{eqnarray}
\nonumber F[N  \to M]  &  =   \sum_l \,   d_l  \,      \< \alpha_l  |  A_l |\alpha_l\> \\
\nonumber & \le   \sum_l  d_l\,     \left(  \sum_{(j,k) \to l}     \sqrt{ \frac {  p_{M,j}  p_{  N,  k}   }{  d_j d_k } }   \,  a_{  jkl}\right)^2  
 \, ,
\end{eqnarray}
where we defined $  a_{  jkl}  :  =  \sqrt{    [A_{l}]_{(j,k)  (j,k)} }  $ and used the positivity of the matrix $A_l$. 
Moreover, we can continue the chain of inequalities as 
\begin{eqnarray*}
 \nonumber  F[  N \to M]  & \le   \sum_l  d_l\,      \left( \max_{ j }  \frac{   p_{M,j}}{d_j^2}   \right)  \,   S  \\
    S   :  =  \sum_l  &  d_l     s_l^2     \qquad   s_l    :  =  \sum_{(j,k) \to l}     \sqrt{ \frac {  p_{  N,  k}   \,  d_j  }{   d_k } }   \,  a_{  jkl}    \,   .
 \end{eqnarray*}
Since from Eq. (\ref{ppp}) we have   $  \max_j \frac{ p_{M,j}}{d_j^2}  =  p_{M,j_{\min}^{(M)}}/d_{j_{\min}^{(M)}}$, the bound becomes 
\begin{eqnarray}
F[N \to M]  \le \frac{p_{M,  j_{\min}^{(M)}}}{d^2_{j^{(M)}_{\min}}} \,    S \, .
\end{eqnarray}

We now   maximize  $S$  under the trace-preservation constraint of Eq. (\ref{ttt}).  In terms of the coefficients $\{a_{jkl}\}$, the  constraint reads  
\begin{eqnarray}\label{nnn}
\sum_{   (j,l)  \to k }    d_l^2     \,   a^2_{jkl}     =     d_k^2  \qquad \forall k  \in \left\{  k_{\min}^{(N)},  \dots,  N/2\right\}  \, . 
\end{eqnarray} 
Now, it is convenient to partition the optimal  coefficients $\{  a_{jkl}\}$  into groups labelled by $k$.  Precisely,  for every $k$ we define the set 
\begin{eqnarray*}
G_k  :  =  \left\{  (j,l) ~:~  (j,l)  \to  k   ,   ~  a_{jkl}  \not  =   0 \right\}.  
\end{eqnarray*}
With this definition, the method of Lagrange multipliers shows that the optimal coefficients $\{a_{jkl}\}$  have the  property 
$$ a_{jkl}   =   \sqrt {d_j}   \,  c_k  \,  \frac{  s_l}{d_l} \qquad  \forall l,\quad \forall (j,l)  \in  G_k \, ,$$
where $c_{k}\ge 0 $  are suitable coefficients.    The normalization condition of Eq. (\ref{nnn}) then becomes 
 \begin{eqnarray}\label{mmm} \sum_{  (j,l )  \in G_k}    d_j     \, c^2_k\,   s_l^2    =  {d_k^2}   \qquad \forall k  \in \left\{  k_{\min}^{(N)},  \dots,  N/2\right\}  \, .    
\end{eqnarray}
Combining the expressions for $S$,  $s_l$,  and for the optimal coefficients $\{a_{jkl}\}$, we then obtain
\begin{eqnarray}
\nonumber S  & \le  \sum_l  d_l  \,  s_l   \left( \sum_{(j,k)  \to l  }  \sqrt{ \frac {  p_{  N,  k}   \,  d_j  }{   d_k } }   \,  a_{  jkl}\right) \\
\nonumber  &  = \sum_k   \sqrt{ \frac {  p_{  N,  k}    }{   d_k } }          \left( \sum_{(j,l)  \in G_k  }  \sqrt{d_j} \, d_l  \, s_l\,  a_{jkl} \right) \\
\nonumber  &  = \sum_k   \sqrt{ \frac {  p_{  N,  k}    }{   d_k } }          \left(\sum_{(j,l)  \in   G_k}    d_j \,   c_{k}  \, s_l^2 \right) \\
\label{zzz}  &  = \sum_{k:  \, c_k  \not = 0}   \sqrt{ \frac {  p_{  N,  k}    }{   d_k } }    \,  \frac{d^2_k}{c_k}    \,  
 \end{eqnarray}
having used Eq. (\ref{mmm}) for the last equality.  Now, the only free variables are the coefficients $\{c_k\}$.  In order to complete the optimization we note that, by definition of $s_l$, we must have 
$$  s_l    =  \sum_{(j,k)\to l}   \sqrt{\frac{  p_{N,k}}{d_k}}  \,  d_j  \,  c_k  \,  \frac{  s_l}{d_l}  \, .$$
which, for $s_l  \not =  0$, implies the constraint  
$$    \sum_{(j,k)  \to l}   \sqrt{\frac{  p_{N,k}}{d_k}}  \,  d_j  \,  c_k  =  d_l   \, .$$
Clearly, for every $c_k  \not  = 0$ there will be at least one value of $l$ such that $s_l  \not  = 0$ and $c_k$ appears in the $l$-th constraint. Let us pick one such value for every $k$---call it  $l(k)$---and define the sets 
$$   H_l  :  =  \left\{   k  ~|~  l(k)  =  l \right\} \, .$$
Now, the r.h.s. of $S$ is upper bounded by the maximum of 
$$  S':=  \sum_{k:  \, c_k  \not = 0}   \sqrt{ \frac {  p_{  N,  k}    }{   d_k } }    \,  \frac{d^2_k}{c_k}     $$
subject to the constraints  
$$     \sum_{k \in  H_l \, , (j,k)  \to l}   \sqrt{\frac{  p_{N,k}}{d_k}}  \,  d_j  \,  c_k  =  d_l   \, .   $$
Maximizing $S'$ under these  constraints we obtain the bound  
$$   S  \le  \sum_{l}   \frac{\left( \sum_{  k\in  H_k}  \sqrt{  p_{N,k}   d_k    \sum_{j:   (j,k) \to l}    d_j    } \right)^2}{d_l}  \, , 
$$  and, using the relation  $\sum_{j:   (j,k) \to l}    d_j       \le d_k d_l$, 
\begin{eqnarray*}
 S  & \le  \sum_{l}   \left( \sum_{  k\in  H_k}  \sqrt{  p_{N,k} }  d_k \right)^2   \le  \left( \sum_{  k =  -N/2}^{N/2}  \sqrt{  p_{N,k} }  d_k \right)^2 
  \, .  
\end{eqnarray*}    
In conclusion, we obtained the  bound $$ F_{clon} [  N  \to M]  \le \frac{p_{M,  j_{\min}^{(M)}}}{d^2_{j^{(M)}_{\min}}}     \,  \left( \sum_{  k=  -N/2}^{N/2}  \sqrt{  p_{N,k}}   d_k \right)^2 \, .$$   
At this point, it is easy to recognize in the r.h.s. the upper bound promised by proposition \ref{prop:twoqubit}: 
First,    using the Schur's lemma and Eqs. (\ref{decompositionmaxent})  and (\ref{aaa}) it is easy show that  the average target state $\rho_{AV}^{(M)}$ is
  \begin{equation*}
\rho_{AV}^{(M)}=\sum_{j=j_{min}^{(M)}}^{M/2}  {p_{M,j}}   ~  \left[   \frac {  I_A^{(j,M)}}  {d_j}    \otimes   \frac {  I_{B}^{(j,M)}}  {d_j}  \otimes    \frac{|  I_{m_j}^{(M)} \>\!\>\<  \! \<     I_{m_j}^{(M)} |}{  m_{j}^{(M)}}  \right]  \, ,
\end{equation*}
where   $ I_A^{(j,M)}$   ( $I_B^{(j,M)}$)   denotes the identity on the representation space $\spc R_A^{(j,M)}$  ($\spc R_B^{(j,M)}$).   Hence, the maximum eigenvalue of  $\rho_{AV}^{(M)}$ is 
\begin{eqnarray*}
\left \|  \rho_{AV}^{(M)}  \right\|_{\infty}   &  =   \max \left\{   \frac{  p_{M,j}}{d_j^2}   ~|~  j  \in \{  j_{min}^{(M)},  \dots, M/2  \}  \right\} \\
 &  =  \frac{ p_{M,j^{(M)}_{\min}}}{  d^2_{j^{(M)}_{\min}}} \, . 
\end{eqnarray*}
Combining this fact with  Eq. (\ref{pns}), we then get to the desired bound 
$ F_{clon} [  N  \to M]  \le   \left \|  \rho_{AV}^{(M)}  \right\|_{\infty}  \,  p_{succ}^{(N)}$.   
 \qed

\section{Fidelity of the MP protocol for two-qubit maximally entangled states}\label{app:MP2qubit}
The fidelity of the protocol  is 
\begin{equation*}
F_K [  N \to M]=\int   \d \hat g  ~     p_N (\hat g|e)   \,   f_{K,M}  (\hat g) \,  ,  
\end{equation*}
with 
$p_N (\hat g|e)    :=   \< \eta_{\hat g} |   \,   \psi^{\otimes N}\,  |\eta_{\hat g}\> $ and
 $f_{K,M} (\hat g)    : =   \left|   \<  \psi|^{\otimes M}    V_{   K,M}    |\psi_{\hat g}\>^{\otimes K}     \right|^2$.
Now, it is easy to see that $p_N (\hat g|0)$ and $ f_{K,M} (\hat g)$ depend only on the rotation angle $\tau$ defined by  the relation 
$  U_{\hat g}  =  \exp \left[   \frac{  \tau   \, \vec n  \cdot \vec \sigma  }{2} \right]  \, ,$
where $\vec n \in \mathbb R^3$ is a unit vector and $\vec n \cdot \vec \sigma  :=  n_x \sigma_x + n_y\sigma_y +  n_z\sigma_z$. 
Precisely, we have 
\begin{eqnarray*}
p_N (\hat g|e)  &=    \left |  \sum_{j=  j_{\min}^{(N)}}^{N/2}  \sqrt{ {p_{N,j}}}     \,  \frac{  \sin [( j+  1/2 )\tau]}{\sin (  \tau/2)} \right|^2  \\
&  =:  p_N (\tau | 0)  
\end{eqnarray*}
and 
\begin{eqnarray*} 
f_{K,M} (\hat g) &=    \left |  \sum_{j=  j_{\min}^{(K)}}^{K/2} \frac{ \sqrt{   p_{K,j}  \,  p_{  M, j}  }  }{d_j}   \,  \frac{  \sin [( j+  1/2 )\tau]}{\sin (  \tau/2)} \right|^2  \\  
   &=:  f_{K,M}  (\tau)  \, .
\end{eqnarray*}
Let us parametrize  the group elements in terms of the rotation angle $\tau$ and of the polar coordinates $ \hat \tau$ and $\hat \psi$ defined by $\vec n  = ( \sin \hat \tau  \cos\hat \psi, \sin \hat \tau \sin \hat \psi, \cos \hat \tau)$.  Recalling the expression of the normalized Haar measure  
$   \d \hat g  =    ({  1}/ {2\pi^2})\,   \left( \sin \frac{\tau} 2\right)^2    \sin\hat \tau \,    \d \tau \d \hat \tau \d \hat \psi $,
 the fidelity can be re-written as 
\begin{eqnarray*}
F_K [  N \to M]=  \frac {  2} {\pi}  \int_{0}^\pi  \d \tau      \left( \sin \frac{\tau} 2\right)^2        p_N (\tau|0)   \,   f_{K,M}  (\tau) \,  ,  
\end{eqnarray*}
 
Now, when $K$ is large, the fidelity function $f_{K,M}  (\tau)$ takes the Gaussian form 
\begin{eqnarray*} 
f_{K,M}  (\tau)  =  \left( \frac{  2 \sqrt{ MK}}{  M+  K} \right)^3   \exp  \left [ -  \frac{   MK \tau^2} { 2 (  M+K)}\right] \left[ 1+  O\left(  \frac 1  {\sqrt{K^{1-\delta}}}\right) \right]   +  \eta_K
\end{eqnarray*} 
 having used the Gaussian approximation of Eq.  (\ref{approx}).   Clearly, when $M$ and $K$ are large compared to $N$,  the slowly varying functions $p_N(\tau|0)$ and $\sin  (\tau/2)$  can be Taylor expanded to the leading order in the integral.  Using this fact, we obtain  
\begin{eqnarray}
\nonumber F_K [  N \to M]  &=    p_N  (0|0 )   \,           \left[   \int_{0}^{\infty}    \frac{  \d \tau}  {2\pi} ~      {\tau^2}   f_{K,M}  (\tau)  \right] \left[ 1+  O\left(  \frac 1  {\sqrt{K^{1-\delta}}}  +  \frac N K\right) \right]   \\
  \nonumber &=    p_{succ}^{(N)}    \,     \sqrt{  \frac{  8} {  \pi  (  M+K)^3}} \left[ 1+  O\left(  \frac 1  {\sqrt{K^{1-\delta}}}  +  \frac NK\right) \right]   \, .    
\end{eqnarray}

\section{Lower bound on the distance between economical and MP channels}\label{app:distance}

{\bf Proof of proposition \ref{distance}.} Consider an economical channel, written as $\map C (\rho)  =  V \rho V^\dag$, and an  MP channel, written as $\widetilde{\map C}  (\rho)  =  \sum_{y\in\mathsf Y}\Tr  [P_y \rho]  ~  \rho_y$. For an arbitrary pure state $|\psi\>\in \spc H_{in}$, the relation between trace distance and fidelity gives \begin{eqnarray}\label{fidelity} \left\|   \map C (\psi)   -\widetilde{ \map   C}   (\psi) \right \|_1  \ge   2 \left(  1  -   \sqrt{ \< \psi  |  V^\dag   \widetilde{\map C}  (\psi)  V |\psi\>   }  \right)  \, . 
\end{eqnarray}   

Moreover, denoting by $\mathsf H$ the convex hull of the states $(\rho_y)_{y\in\mathsf Y}$ we have   
\begin{eqnarray*}
\< \psi  |  V^\dag   \widetilde{\map C}  (\psi)  V |\psi\>  &\le  \max_{\sigma\in\mathsf H}  ~   \< \psi  |  V^\dag  \sigma V |\psi\>   
\end{eqnarray*} 
 and, minimizing over $ |\psi\> $,        
$\min_{|\psi\> \in \spc H_{in}} ~  \< \psi  |  V^\dag   \widetilde{\map C}  (\psi)  V |\psi\>  \le  \min_{\rho  \in  \mathsf S }   \max_{\sigma \in \mathsf H}~  \Tr  [  V \rho V^\dag \sigma]$ , 
where 
$\mathsf S $ denotes the set of all quantum states on $\spc H_{in}$.  
Now, von Neumann's minimax's theorem allows us to  exchange the minimum and the maximum in the r.h.s, thus obtaining 
\begin{eqnarray*}
\min_{|\psi\> \in \spc H_{in}} ~  \< \psi  |  V^\dag   \widetilde{\map C}  (\psi)  V |\psi\>  &\le  \max_{\sigma \in \mathsf H}   \min_{\rho  \in    \mathsf S }   \Tr  [  V \rho V^\dag \sigma]  \\
&  =  \max_{\sigma \in \mathsf H} ~  \tilde \sigma_{\min}
\end{eqnarray*}
where $\tilde \sigma_{\min}$ denotes the minimum  eigenvalue of $\tilde \sigma:  =  P  \sigma  P$, $P$ being the projector on the subspace $V\spc H_{in}$.  Since, $\tilde \sigma$ is a non-negative matrix with $\Tr [\tilde \sigma  ]\le 1$ and with rank upper bounded by $d_{in}$, its minimum eigenvalue is upper bounded by $d_{in}^{-1}$.  
  Hence, we obtained 
$\min_{|\psi\>\in\spc H_{in} } ~  \< \psi  |  V^\dag   \widetilde{\map C}  (\psi)  V |\psi\>  \le  d_{in}^{-1} ,$
 and, therefore, 
 $\left\|  \map C  -  \widetilde{\map C} \right\|_1     \ge     2    \left(   1  -  \sqrt{ 	d^{-1}_{in} } \right)$.   
$\blacksquare$ 

\end{document}